\documentclass[%
groupedaddress,
 amsmath,amssymb,
 aps,
prb,
]{revtex4-2}

\usepackage{graphicx}
\usepackage{dcolumn}
\usepackage{bm}
\usepackage{siunitx} 
\usepackage{hyperref}
\usepackage{graphicx}
\usepackage{epstopdf}
\usepackage{rotating}
\usepackage{array}
\usepackage{color}
\usepackage{slashed}
\usepackage{multirow}
\usepackage{amsmath}
\usepackage[makeroom]{cancel}
\usepackage{tabularx}

\begin{document}


\title{\Large The  $Z_3$ soft breaking in the I(2+1)HDM\\ and its probes at present and future colliders}

\author{D. Hern\'andez-Otero}
\email[]{danielah@ifuap.buap.mx}
\affiliation{Instituto de F\'isica, Benem\'erita Universidad Aut\'onoma de Puebla,  Apdo. Postal J-48, C.P. 72570 Puebla, Puebla, M\'exico}

\author{J. Hern\'andez-S\'anchez}
\email[]{jaime.hernandez@correo.buap.mx}
\affiliation{Facultad de Ciencias de la Electr\'onica, Benem\'erita Universidad Aut\'onoma de Puebla, Apartado Postal J-48, 72570 Puebla, M\'exico}
\affiliation{Dual CP  Institute of High Energy Physics, Puebla,  M\'exico}

\author{S. ~Moretti}
\email[]{S.Moretti@soton.ac.uk}
\email[]{stefano.moretti@physics.uu.se}
\affiliation{School of Physics and Astronomy, University of Southampton, Southampton, SO17 1BJ, United Kingdom}
\affiliation{Department of Physics and Astronomy, Uppsala University, Box 516, SE-751 20 Uppsala, Sweden}

\author{T.~Shindou}
\email[]{shindou@cc.kogakuin.ac.jp}
\affiliation{Division of Liberal-Arts, Kogakuin University,  2665-1 Nakano-machi, Hachioji, Tokyo, 192-0015, Japan}

\date{\today}

\begin{abstract}
A $ Z_3 $ symmetric 3-Higgs Doublet Model (3HDM) with two inert doublets and one active doublet (that plays the role of the Higgs doublet),  the so-called I(2+1)HDM, is studied. We discuss the behaviour of this 3HDM realisation when one allows for a $ Z_3 $ soft breaking term.  Such a symmetry enables the presence of a two-component Dark Matter (DM) scenario in 
the form of  ``Hermaphrodite DM",  where the two inert candidates have opposite CP parity and are protected by this discrete symmetry from decaying into Standard Model (SM) particles. Furthermore, the two DM states are potentially distinguishable from each other as they cannot be subsumed into a complex field, having different masses and gauge couplings. With this
in mind, we study differential spectra with a distinctive shape from which the existence of two different DM component distributions could be easily inferred. We prove this to be possible
at  the Large Hadron Collider (LHC) via the $q\bar{q} \to 2l+H_1H_1$ and $q\bar{q}  \to 2l+A_1A_1$ processes as well as at a future electron-positron machine via  the $e^{+}e^{-} \to 2l+H_1H_1$ and $e^{+}e^{-} \to 2l+A_1A_1$ channels, where $l=e,\mu$.
\end{abstract}

\maketitle


\section{\label{introduction} Introduction}

A Higgs boson was discovered at the Large Hadron Collider (LHC) in July 2012 \cite{ATLAS:2012yve, CMS:2012qbp} 
and it has been shown that its nature is consistent with the one of the Standard Model (SM), 
which contains only one SU(2) doublet field.
However, no compelling principle has ever been put forward that constrains  the Higgs sector responsible for Electro-Weak Symmetry Breaking (EWSB) and mass generation to be
the one of the SM. 
In particular, there is no reason which forbids introducing new fields into the (pseudo)scalar sector of the underlying theory. 

There are many possibilities to extend the Higgs sector of the SM. As Nature seems to prefer doublet (pseudo)scalar fields, one could well restrict oneself to extensions of the SM that
only include such representations. Indeed, 
 the ensuing $N$-Higgs Doublet Model (NHDM) is a simple and attractive example, 
where the Higgs sector contains $N$ such fields.
The extension with doublet (pseudo)scalar fields also keeps $\rho=1$ at the tree level.
However, NHDMs generally lead to dangerous Flavour Changing Neutral Currents (FCNCs).
In order to suppress these, a discrete symmetry is often utilised. 
For example, in the 2HDM, which is well studied in the literature, a softly broken $Z_2$ symmetry is usually introduced.
Under this $Z_2$ symmetry, one doublet is odd, and the other doublet is even.
Depending on the $Z_2$ parity assignment for the SM fermions, 2HDMs are then classified into four types \cite{Barger:1989fj,Grossman:1994jb,Aoki:2009ha}.

There are reasons to consider this kind of SM extension, as there are some problems that the SM cannot explain, calling for new physics. For example, within the SM, there is no viable candidate for Dark Matter (DM), there is no successful mechanism for baryogenesis, there is no dynamics that explains the smallness of the neutrino masses and so on.
One should then expect that Beyond the SM (BSM) scenarios can solve these problems, and specifying the Higgs sector provides a vital clue to explore such new physics. NHDMs can, in particular, afford one with viable DM candidates.

For example, the so-called inert 2HDM, also acronym IDM, provides one such DM candidate. 
In this model, even parity is assigned to the SM fermions, 
there is no softly broken term acting on the $Z_2$ symmetry in the Higgs potential, and only the $Z_2$ even doublet gets a Vacuum Expectation Value (VEV) \cite{Deshpande:1977rw}.
Since the $Z_2$ is kept unbroken, the lightest $Z_2$ odd scalar is, therefore, stable, and it can be a DM if the particle is neutral. 

The inert sector of an NHDM with more than two doublets also provides DM candidates.
In the case of a 3-Higgs Doublet Model (3HDM), there are many possibilities to impose a discrete symmetry and trigger its breaking patterns. 
DM properties in an unbroken $Z_3$ symmetry model that lead to two inert doublets are discussed in Ref.~\cite{Aranda:2019vda}.
If the CP symmetry is unbroken in the inert sector,  the lightest $Z_3$ charged particles in the CP-odd and CP-even sectors are individually stable. 
Thus two DM candidates with opposite CP charges are provided.  This realisation is called a ``Hermaphrodite DM" scenario. 

In this paper, we analyse collider signals of this scenario. 
Since the two DM particles interact with the $Z$ and $W$ gauge bosons and the SM Higgs boson, 
they are produced at collider experiments such as the LHC and/or a future $e^+e^-$ collider. We will prove that
it should be possible to reveal both these two DM candidates by isolating phenomenological properties of processes leading to the same final state, proceeding through the two 
different states and pointing to their simultaneous presence. 

This paper is organised as follows.
Sec. II introduces the model with its Hermaphrodite DM scenario.
In Sec. III, we discuss the parameters used in our analysis. 
We show our numerical results in Sec. IV. 
Finally, we give our conclusions in Sec. V. 

\section{The  model \label{model}}
\subsection{Lagrangian}
In this paper, 
we consider an extended Higgs sector with three Higgs doublets $\phi_i~(i=1,2,3)$.
We impose a $Z_3$ symmetry under which the three doublets transform as 
\begin{equation}
	\phi_1^{}\to \omega \phi_1^{}\;,\quad 
	\phi_2^{}\to \omega^2 \phi_2^{}\;, \quad 
	\phi_3^{}\to \phi_3^{}\;,
\end{equation}
with $\omega$ being a complex cubic root of unity, \textit{i.e.}, $\omega=e^{2\pi i/3}$.
The symmetric Higgs potential is given by 
\begin{equation}
	V=V_0+V_{Z_3}
\end{equation}
where $V_0$ is an invariant part under any phase rotation given by 
\begin{align}
	V_0 =& - \mu^2_{1} (\phi_1^\dagger \phi_1) -\mu^2_2 (\phi_2^\dagger \phi_2) - \mu^2_3(\phi_3^\dagger \phi_3) \\
	&+ \lambda_{11} (\phi_1^\dagger \phi_1)^2+ \lambda_{22} (\phi_2^\dagger \phi_2)^2  + \lambda_{33} (\phi_3^\dagger \phi_3)^2 \nonumber\\
	& + \lambda_{12}  (\phi_1^\dagger \phi_1)(\phi_2^\dagger \phi_2)
	+ \lambda_{23}  (\phi_2^\dagger \phi_2)(\phi_3^\dagger \phi_3) + \lambda_{31} (\phi_3^\dagger \phi_3)(\phi_1^\dagger \phi_1) \nonumber\\
	& + \lambda'_{12} (\phi_1^\dagger \phi_2)(\phi_2^\dagger \phi_1) 
	+ \lambda'_{23} (\phi_2^\dagger \phi_3)(\phi_3^\dagger \phi_2) + \lambda'_{31} (\phi_3^\dagger \phi_1)(\phi_1^\dagger \phi_3) \nonumber
	\end{align}
and $V_{Z_3}$ is a collection of extra terms ensuring the 
$Z_3$ symmetry given by 
\begin{equation}
	V_{Z_3} = \lambda_1(\phi_2^\dagger\phi_1)(\phi_3^\dagger\phi_1) + \lambda_2(\phi_1^\dagger\phi_2)(\phi_3^\dagger\phi_2) + \lambda_3(\phi_1^\dagger\phi_3)(\phi_2^\dagger\phi_3)  + \text{h.c.}
	\label{Z_3-3HDM}
\end{equation}

We adopt an ansatz that only $\phi_3$ has a VEV.
With this assumption, the EW symmetry is broken by $\langle \phi_3\rangle$ 
while the $Z_3$ symmetry is (initially) unbroken. 
A physical component in the $Z_3$ singlet field $\phi_3$ behaves like the SM Higgs boson, so we describe it as such using the label $h$.
Also, all SM particles have a $ Z_3 $ zero charge, so that only $ \phi_3 $ will couple to fermions. 
The Yukawa Lagrangian is given by
\begin{eqnarray}
	\mathcal{L}_{Y} &=& \Gamma^u_{mn} \bar{q}_{m,L} \tilde{\phi}_3 u_{n,R} + \Gamma^d_{mn} \bar{q}_{m,L} \phi_3 d_{n,R} \nonumber\\
	&& +  \Gamma^e_{mn} \bar{l}_{m,L} \phi_3 e_{n,R} + \Gamma^{\nu}_{mn} \bar{l}_{m,L} \tilde{\phi}_3 {\nu}_{n,R} + \text{h.c.}
\end{eqnarray}
Thanks to the $Z_3$ symmetry,  the lightest components of $\phi_1$ and $\phi_2$ can be stable, and they both are DM candidates.
Given that $\phi_1$ and $\phi_2$ are inert, 
this model is termed I(2+1)HDM \cite{Ivanov:2011ae,2021symm}.
Since the CP symmetry is also kept in the potential,  the combination of the CP and $Z_3$ symmetries  
predicts that these two DM candidates are such that one is CP-even and the other is CP-odd. However, evidently not being the real and imaginary part
of a complex field (as it will be clear below),  such a two-component DM is aptly named Hermaphrodite DM  \cite{2021symm}.

As we will discuss later, the model with the softly broken term, only affecting the inert sector,   
\begin{equation}
	V_{\slashed{Z_3}}= -\mu_{12}^2(\phi_1^{\dagger}\phi_2)+\text{h.c.} 
\end{equation}
is the one interesting from the phenomenological point of view, though. 
 In fact, to realise proper EWSB, the parameter $ \mu_{12}^2 $ must be small, thus allowing for a $Z_3$ symmetry soft breaking. As a consequence, the stability of the DM candidates is not affected.

\subsection{The physical eigenstates}
The scalar potential acquires a minimum at the point
\begin{equation}
	\phi_1= 
	\left( \begin{array}{c}{\scriptstyle{H^{0+}_1}}\\ \frac{H^0_1+iA^0_1}{\sqrt{2}} \end{array}\right), \qquad \phi_2= 
	\left( \begin{array}{c}{\scriptstyle{H^{0+}_2}}\\ \frac{H^0_2+iA^0_2}{\sqrt{2}} \end{array}\right), \qquad \phi_3= 
	\left( \begin{array}{c}{\scriptstyle{H^{0+}_3}}\\ \frac{H_3^{0}+v+iA^0_3}{\sqrt{2}} \end{array}\right),
	\label{fields}
\end{equation}
where $v^2=\mu_{3}^{2}/\lambda_{33}$. 
Expanding the potential around the vacuum point we obtain the mass spectrum. In the model that allows a soft breaking of the $Z_3$ symmetry we have the following.
\begin{itemize}
	\item The neutral sector, CP-even scalars:
	\begin{eqnarray}
		&& \textbf{h} = H_3^{0} : \quad m^2_{h}= 2\mu_3^2 = 2 \lambda_{33} v^2.\\[1mm]
		&& \textbf{H}_1 = \cos\theta_h H^0_{1}+ \sin\theta_h H^0_{2}  
		\nonumber\\[1mm]
		&& \hspace{1cm}	
		m^2_{H_1}=  (-\mu^2_1 + \Lambda_{1})\cos^2\theta_h + (- \mu^2_2 + \Lambda_{2}) \sin^2\theta_h - (2\mu^2_{12} - \lambda_3 v^2) \sin\theta_h \cos\theta_h. 
		\nonumber\\[1mm]
		&& \textbf{H}_2 = -\sin\theta_h H^0_{1}+ \cos\theta_h H^0_{2} 
		\nonumber \nonumber\\
		&& \hspace{1cm}	
		m^2_{H_2}=  (-\mu^2_1 + \Lambda_{1})\sin^2\theta_h + (- \mu^2_2 + \Lambda_{2}) \cos^2\theta_h + (2\mu^2_{12} - \lambda_3 v^2) \sin\theta_h \cos\theta_h. 
		\nonumber
	\end{eqnarray}
	\item The neutral sector, CP-odd scalars:
	\begin{eqnarray}
		&& \textbf{A}_1 = \cos\theta_a A^0_{1}+ \sin\theta_a A^0_{2} \\[1mm]
		&& \hspace{1cm}	
		m^2_{A_1}= (-\mu^2_1 + \Lambda_{1})\cos^2\theta_a + (- \mu^2_2 + \Lambda_{2}) \sin^2\theta_a - (2\mu^2_{12} + \lambda_3 v^2) \sin\theta_a \cos\theta_a. 
		\nonumber\\[1mm]
		&& \textbf{A}_2 = -\sin\theta_a A^0_{1}+ \cos\theta_a A^0_{2}	\nonumber\\[1mm] && \hspace{1cm}		
		m^2_{A_2}= (-\mu^2_1 + \Lambda_{1})\sin^2\theta_a + (- \mu^2_2 + \Lambda_{2}) \cos^2\theta_a + (2\mu^2_{12} + \lambda_3 v^2) \sin\theta_a \cos\theta_a. \nonumber
	\end{eqnarray}
	\item The charged sector:
	\begin{eqnarray}
		&& \textbf{H}^\pm_1 = \cos\theta_c H^{0\pm}_{1} + \sin\theta_c H^{0\pm}_{2}\\[1mm]
		&& \hspace{1cm}	
		m^2_{H^\pm_1}= (-\mu^2_1 + \frac{1}{2}\lambda_{31}v^2)\cos^2\theta_c + (- \mu^2_2 + \frac{1}{2}\lambda_{23}v^2) \sin^2\theta_c - 2\mu^2_{12}\sin\theta_c \cos\theta_c. 
		\nonumber\\[1mm]
		&& \textbf{H}^\pm_2 = -\sin\theta_c H^{0\pm}_{1} + \cos\theta_c H^{0\pm}_{2}
		\nonumber\\[1mm]
		&& \hspace{1cm}	
		m^2_{H^\pm_2}= (-\mu^2_1 +\frac{1}{2}\lambda_{31}v^2)\sin^2\theta_c + (- \mu^2_2 + \frac{1}{2}\lambda_{23}v^2) \cos^2\theta_c + 2\mu^2_{12}\sin\theta_c \cos\theta_c. \nonumber
	\end{eqnarray}
\end{itemize}
Here, $\Lambda_1=(\lambda_{31}+\lambda^{\prime}_{31})v^2/2$ and $\Lambda_1=(\lambda_{23}+\lambda^{\prime}_{23})v^2/2$. The angles $\theta_h, \theta_a, \theta_c$ are the mixing angles for the scalar, pseudoscalar an charged mass-squared matrices, respectively, such that:
\begin{eqnarray}
	&& \tan 2\theta_c=\frac{4\mu_{12}^2}{2\mu_1^2-\lambda_{31}v^2-2\mu_2^2+ \lambda_{23}v^2}  =\varepsilon_c, \nonumber\\[1mm]
	&&\tan 2\theta_h=\frac{-\lambda_{3}v^2+2\mu_{12}^2}{\mu_1^2-\Lambda_1-\mu_2^2+ \Lambda_2} =- \frac{\lambda_3 v^2 }{\mu^2_1 - \Lambda_1 - \mu^2_2 + \Lambda_2}  + \varepsilon_h, \nonumber\\[1mm]&& \tan 2\theta_a=\frac{\lambda_{3}v^2+2\mu_{12}^2}{\mu_1^2-\Lambda_1-\mu_2^2+ \Lambda_2}=\frac{\lambda_3 v^2 }{\mu^2_1 - \Lambda_1 - \mu^2_2 + \Lambda_2}  + \varepsilon_h,
\end{eqnarray}
in terms of the small parameter $\varepsilon_h=\frac{2\mu_{12}^{2} }{\mu^2_1 - \Lambda_1 - \mu^2_2 + \Lambda_2}  $. From the expressions above we can extract the relation $\tan 2\theta_a = -\tan 2\theta_h + 2\varepsilon_h$. Since the $Z_3$ is softly broken via the small $\mu_{12}^2$ term, then we can write  the rotation angles as
\begin{equation}
	\theta_a=-\theta_h+\epsilon_h \quad \text{and} \quad \theta_c=\epsilon_c.
\end{equation}
The masses squared for the  (pseudo)scalars are
\begin{eqnarray}
	m_{H_1}^2&=&(-\mu_1^2+\Lambda_1)\cos^2\theta_h+(-\mu_2^2+\Lambda_2)\sin^2\theta_h-\frac{1}{2}(2\mu_{12}^2-\lambda_{3}v^2)\sin2\theta_h,\\[1mm]
	m_{A_1}^2&=&m_{H_1}^2+\left[2\mu_{12}^2+\epsilon_h(-\mu_1^2+\Lambda_1+\mu_{2}^2-\Lambda_2)\right]\sin2\theta_h-\epsilon_h(2\mu_{12}^2+\lambda_3v^2)\cos2\theta_h,\nonumber\\[1mm]
	m_{H_2}^2&=&(-\mu_1^2+\Lambda_1)\sin^2\theta_h+(-\mu_2^2+\Lambda_2)\cos^2\theta_h+\frac{1}{2}(2\mu_{12}^2-\lambda_{3}v^2)\sin2\theta_h,\nonumber\\[1mm]
	m_{A_2}^2&=&m_{H_2}^2-\left[2\mu_{12}^2+\epsilon_h(-\mu_1^2+\Lambda_1+\mu_{2}^2-\Lambda_2)\right]+\epsilon_h(2\mu_{12}^2+\lambda_3v^2)\cos2\theta_h.\nonumber
\end{eqnarray}

As we have mentioned above, the CP symmetry makes $H_1$ and $A_1$ stable, so they can both be DM.
After EWSB, 
we have the vertex $ H_1A_1Z \propto \cos2\theta_h+\epsilon_h\sin2\theta_h$, which 
leads to a too large cross-section for DM scattering
off nuclei and thus direct detection immediately 
excludes the scenario. 
In order to avoid it, the coupling constant of this vertex
should be significantly suppressed.  When we choose $\theta_h=\pi/4$, 
the vertex is proportional to $\epsilon_h$, and 
it is thus expected to be small. 
Note that $\epsilon_h=0$ is satisfied in the $Z_3$ symmetric
limit and two DM candidates, $H_1$ and $A_1$ are degenerate in mass, $m_{H_1}=m_{A_1}$.

\section{Parameters for the analysis}
\subsection*{The input parameters}
As mentioned, in order to avoid the model being ruled out by direct detection bounds, we will consider the limit $ \theta_h = \pi / 4 $ where the masses squared can be written as follows: 
\begin{eqnarray}
	m_{H_1}^2&=&\frac{1}{2}(-\mu_1^2+\Lambda_1)+\frac{1}{2}(-\mu_2^2+\Lambda_2)-\frac{1}{2}(2\mu_{12}^2-\lambda_{3}v^2),\nonumber\\[1mm]
	m_{A_1}^2&=&m_{H_1}^2+\left[2\mu_{12}^2+\epsilon_h(-\mu_1^2+\Lambda_1+\mu_{2}^2-\Lambda_2)\right]\label{m1}\\[1mm]
	m_{H_2}^2&=&\frac{1}{2}(-\mu_1^2+\Lambda_1)+\frac{1}{2}(-\mu_2^2+\Lambda_2)+\frac{1}{2}(2\mu_{12}^2-\lambda_{3}v^2),\nonumber\\[1mm]
	m_{A_2}^2&=&m_{H_2}^2-\left[2\mu_{12}^2+\epsilon_h(-\mu_1^2+\Lambda_1+\mu_{2}^2-\Lambda_2)\right],\label{m2}\\[1mm]
	m_{H_1^{\pm}}^2&=&-\mu_1^2+\frac{v^2}{2}\lambda_{31}-2\mu_{12}^2 \epsilon_c,\nonumber\\[1mm]
	m_{H_2^{\pm}}^2&=&-\mu_2^2+\frac{v^2}{2}\lambda_{23}+2\mu_{12}^2 \epsilon_c. \label{m3}
\end{eqnarray}
The input parameters $\lambda_{23}, \lambda_{13}, \lambda_{23}^{\prime}, \lambda_{31}^{\prime},  \mu_1^2$ and $\mu_2^2$ can be rewritten in terms of the physical observables $m_{H_1}, m_{H_2}, m_{H_1}^{\pm}, m_{H_2}^{\pm}, \lambda_{1}$ and  $\lambda_{2}$. Introducing
\begin{eqnarray}
	&&\Delta_h=m_{A_1}-m_{H_1}\\ &&\Delta_c=m_{H_1^{\pm}}-m_{H_1},\\
	&&\delta_c=m_{H_2^{\pm}}-m_{H_1^\pm},\\
	&&\Delta_n=m_{A_2}-m_{H_1} >50~ {\rm GeV},  \label{Dn} \\ &&\Delta_n'=m_{H_2}-m_{A_1}, \\
	&&m_{H_2}^2\approx m_{A_2}^2+2\mu_{12}^2,
	\label{Delm}
\end{eqnarray}
\begin{eqnarray}
	g_1&=&\frac{g_{hH_1H_1}}{v}=\frac{1}{2}(\lambda_{23}+2\lambda_{3}+\lambda_{31}+\lambda_{23}^{\prime}+\lambda_{31}^{\prime}),\nonumber\\ g_2&=&\frac{g_{hH_1H_2}}{v}=\frac{1}{2}(\lambda_{23}-\lambda_{31}+\lambda_{23}^{\prime}-\lambda_{31}^{\prime}),
	\label{g12}
\end{eqnarray}
 where $g_{hH_1H_1}$ and $g_{hH_1H_2}$ are the coefficients of the vertices $hH_1H_1$ and $hH_1H_2$, respectively, for $\epsilon_h\sim 0$, the Lagrangian parameters in terms of the observables  reduce to:
\begin{eqnarray}
	\lambda_{23}&=&\frac{1}{v^2}\left((g1+g2)v^2-2m_{H_1}^{2}+2m_{H_2^{\pm}}\right) -\frac{2\mu_{12}^{2}}{v^2},\label{par1}\\[1mm]
	\lambda_{31}&=&\frac{1}{v^2}\left((g1-g2)v^2-2m_{H_1}^{2}+2m_{H_1^{\pm}}\right)+\frac{2\mu_{12}^{2}}{v^2},\nonumber\\[1mm]
	\lambda_{23}^{\prime}&=&\frac{1}{v^2}\left(m_{H_1}^2+m_{H_2}^2-2m_{H_2^{\pm}}^{2}\right),\nonumber\\[1mm]
	\lambda_{31}^{ \prime}&=& \frac{1}{v^2}\left(m_{H_1}^2+m_{H_2}^2-2m_{H_1^{\pm}}^2\right),\nonumber\\[1mm]
	\lambda_3&=&\frac{1}{v^2}\left(m_{H_1}^{2}-m_{H_2}^2+2\mu_{12}^2 \right),\nonumber\\[1mm]
	\mu_1^2&=&\frac{1}{2}\left( (g_1 -g_2) v^2-2m_{H_1}^2\right)+\frac{2\mu_{12}^{2}}{v^2},\nonumber\\[1mm]
	\mu_2^2&=&\frac{1}{2}\left( (g_1 +g_2) v^2-2m_{H_1}^2\right)-\frac{2\mu_{12}^{2}}{v^2}.\nonumber
\end{eqnarray}



\subsection*{Constraints on the model parameters}\label{constraints}
All the Benchmark Points (BPs) considered in this study  agree with the latest theoretical and experimental constraints that are applicable to the model, which are described in detail in \cite{Keus:2014isa,2021symm}. For convenience, we recap these here.
As $\phi_3$ is identified with the SM Higgs doublet, 
$\mu_3$ and $ \lambda_{33}$ are Higgs field parameters and can be written in terms of the mass of the Higgs boson. We use the value $m_h=125$~GeV for the latter, so that 
\begin{equation} 
	m^2_h = 2\mu^2_3 = 2\lambda_{33} v^2.
\end{equation}
In agreement with perturbativity bounds and unitary conditions, we take the absolute values $|\lambda_i |\leq 3\pi$.
For the potential to be bounded from below, the following conditions are required 
\begin{eqnarray}
	&& \bullet ~ \lambda_{11}, \,\lambda_{22}, \,\lambda_{33} \geq 0, \\[1mm]
	&& \bullet ~  \lambda_{12} + \lambda'_{12}  + \sqrt{\lambda_{11}\lambda_{22}} \geq 0, \nonumber\\[1mm]
	&& \bullet ~ \lambda_{23} + \lambda'_{23} + \sqrt{\lambda_{22}\lambda_{33}} \geq 0,\nonumber\\[1mm]
	&& \bullet ~ \lambda_{31} + \lambda'_{31} + \sqrt{\lambda_{33}\lambda_{11}} \geq 0,\nonumber\\[1mm]
	&& \bullet ~ 
	\sqrt{\lambda_{11}\lambda_{22}\lambda_{33}} + (\lambda_{12} + \lambda'_{12}) \sqrt{\lambda_{33}} 
	+ (\lambda_{31} + \lambda'_{31}) \sqrt{\lambda_{22}} 	+ (\lambda_{23} + \lambda'_{23}) \sqrt{\lambda_{11}} \nonumber\\[1mm]
	&&
	+\sqrt{2 (\lambda_{12} + \lambda'_{12}  + \sqrt{\lambda_{11}\lambda_{22}})(\lambda_{23} + \lambda'_{23} + \sqrt{\lambda_{22}\lambda_{33}})(\lambda_{31} + \lambda'_{31} + \sqrt{\lambda_{33}\lambda_{11}} )} \geq 0.
	\nonumber
\end{eqnarray}
For the $V_{Z_3}$ term not to dominate the behaviour of $V$, we also require the parameters of the $V_{Z_3}$ part to be smaller than the parameters of the $V_0$ part:
\begin{equation}
	|\lambda_1|, |\lambda_2|, |\lambda_3| < |\lambda_{ii}|, |\lambda_{ij}|, |\lambda'_{ij}| , \quad i\neq j = 1,2,3.
\end{equation}
Finally, as intimated, in $V_{\cancel{Z_3}} $, the parameter $\mu_{12}^2$ must be small. 

In our numerical studies, we have taken into account the following limits.
 In agreement with measurements done at LEP 	\cite{Cao:2007rm,Lundstrom:2008ai},  the limit on invisible decays of $ Z $ and $ W ^ \pm $ gauge bosons imply: 

\begin{equation}
m_{H_i^\pm} + m_{H_i,A_i} > m_{W^\pm} , \quad  m_{H_i} + m_{A_i} > m_Z, \quad 2m_{H_i^\pm} > m_Z . \end{equation}
Also,  the lower limit for the mass of the charged scalars  is $m_{H^\pm_i} > 70-90 ~\mbox{GeV}.$
Furthermore, searches for charginos and neutralinos at LEP have been translated into limits on region of masses in the I(1+1)HDM \cite{Lundstrom:2008ai}, simultaneously requiring ($i=1,2$)
\begin{equation}
m_{H_i} \leq 80 ~ \mbox{GeV},  \quad m_{A_i} \leq 100 ~\mbox{GeV}\quad \mbox{and} \quad m_{A_i} - m_{H_i}  \geq  8~\mbox{GeV},  \nonumber
\end{equation}
otherwise a visible di-lepton or di-jet signal could appear.

The decay width of the SM-like Higgs boson into a pair of the inert scalars with $m_{S_i} < m_h/2$ is given by
\begin{equation}
	\Gamma (h\to S_iS_j) = \frac{g^2_{hS_iS_j} v^2}{32\pi m_h^3}
	\biggl[ \biggl(m_h^2-(m_{S_i}+m_{S_j})^2 \biggr)
	\biggl(m_h^2-(m_{S_i}-m_{S_j})^2 \biggr)
	\biggr]^{1/2},
	\label{Eq:Gamma_inv}
\end{equation}

where $S_i,S_j = H_1,A_1$, the coefficient $g_{hS_iS_j}\, v$ corresponds to the $hS_iS_j$ term in the Lagrangian and $m_{S_i}(m_{S_j})$ is the mass of the corresponding neutral inert particle $S_i (S_j)$. From the ATLAS experiment, it is possible to estimate the limit the SM-like Higgs boson invisible Branching Ratios (BR) as $\text{BR}(h\to \text{invisibles}) < 0.08-0.15.$ \cite{Aaboud:2019rtt} . Therefore, we have strong constraints on the {Higgs-DM} coupling. For our scenarios this BR is: 
\begin{equation}
	\text{BR}(h\to \text{invisibles}) = \frac{\Sigma_{i}\Gamma(h\to S_iS_i)}{\Gamma^{\text{SM}}_h + \Sigma_{i}\Gamma(h\to S_iS_i)},
	\label{Eq:BRinv}
\end{equation}
where $S_i =H_1, \, A_1$.  Due to the constraints coming from $h\to\gamma\gamma$,  the parameters and the inert masses in our model  are in agreement with experiments as the one presented in \cite{Cordero-Cid:2018man}.
%

Considering DM constraints, the prediction of the total relic density due to the presence of both $H_1$ and  $A_1$ is given by $\Omega_{\rm DM}h^2=\Omega_{H_1}h^2+\Omega_{A_1}h^2$. The relic density constraint, according to the last measurements from the Planck experiment \cite{plank2020}, is  
\begin{equation*}
\Omega_{\rm DM}h^2 = 0.120 \pm 0.001.
\end{equation*}

The direct detection results of DM measured in XENON1T \cite{Aprile:2018dbl} were then  used. Finally,  the indirect detection results of FermiLAT \cite {Karwin:2016tsw} were 
also adopted (these strongly restrict the annihilation of DM in the final states $ b \bar b $ and $ \tau^+ \tau^- $ \cite{2021symm}).
\subsection*{Parameter scan}
As we discussed previously, for  the $H_1$ and $A_1$ particles to qualify as viable DM candidates, the  $ H_1A_1Z \propto \cos2\theta_h+\epsilon_h\sin2\theta_h$ vertex 
must vanish. Therefore, $\theta_h = \pi/4$ is the only acceptable value in the $0 \leq \theta_h < \pi$ range and $\epsilon_h \propto \mu_{12}^2$ must be small for the model to qualify as a viable DM framework.  Specifically, the constrains on the parameters,  according to the invisible Higgs decay rates, requires that the coefficient of the $hH_1H_2$ vertex must be  $-0.029 \leq g_1 \leq 0.029$.

\begin{center}
	\begin{figure}[!t]
		\centering
		\includegraphics[scale=0.32315]{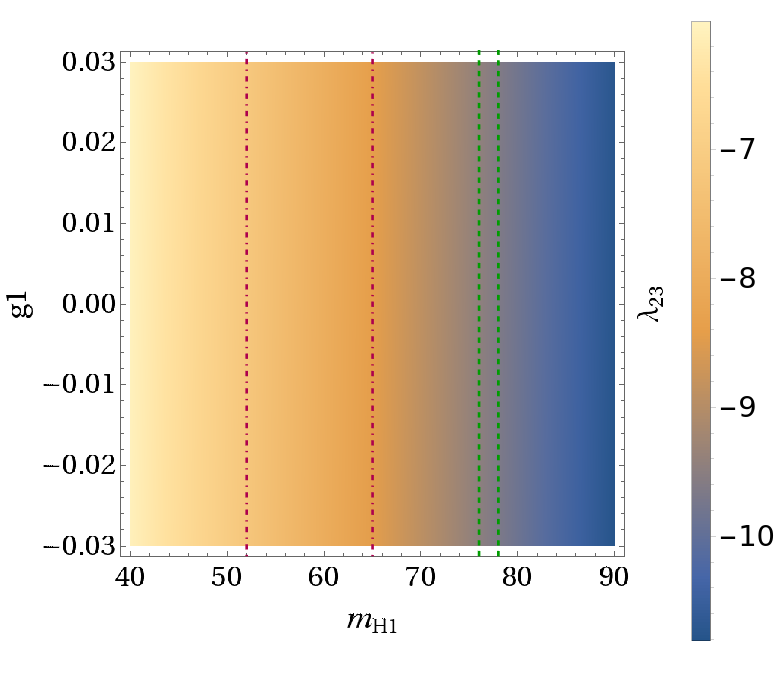}
		\includegraphics[scale=0.32315]{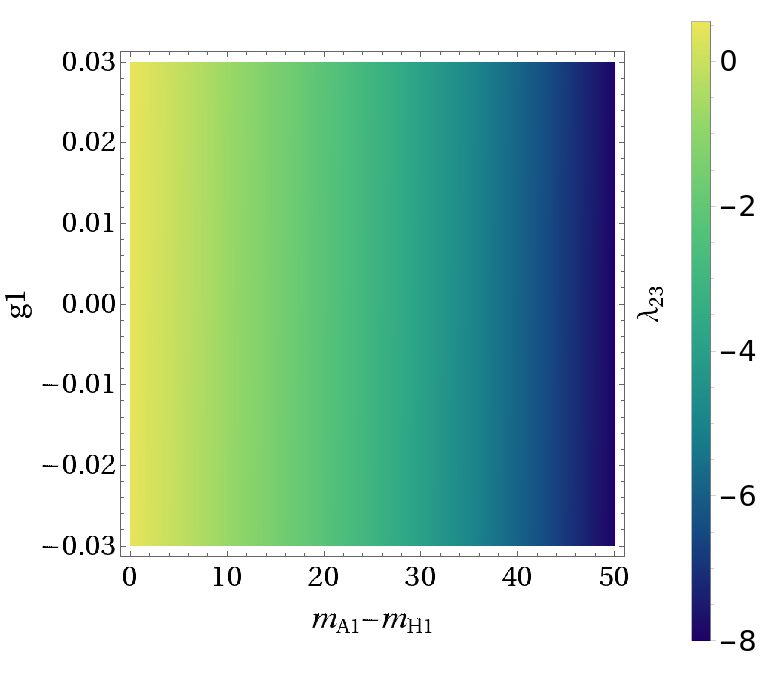}
		\includegraphics[scale=0.32315]{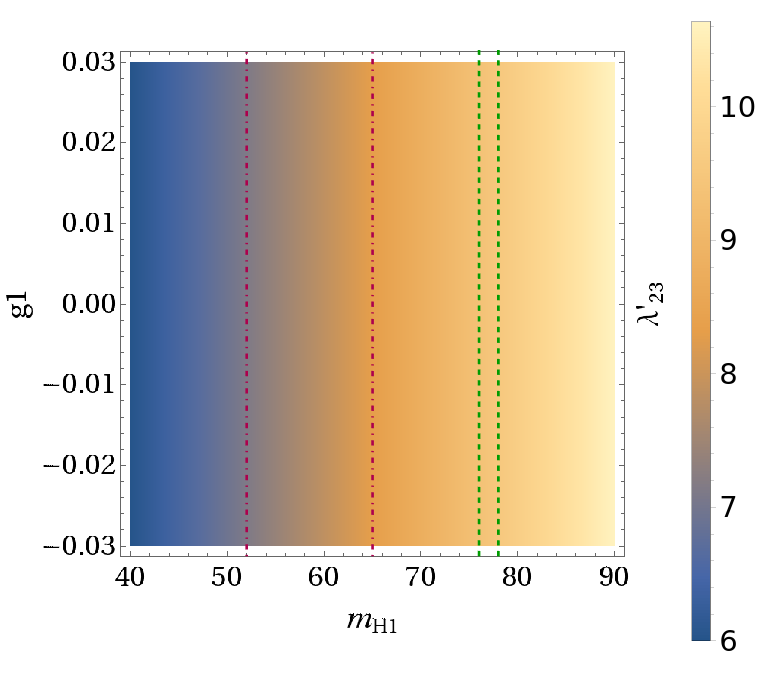}
		\includegraphics[scale=0.32315]{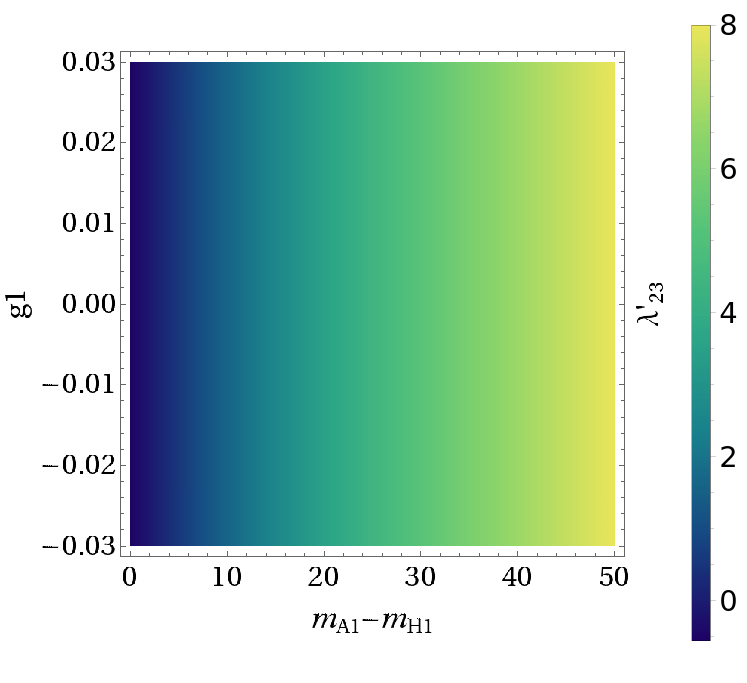}
		\includegraphics[scale=0.32315]{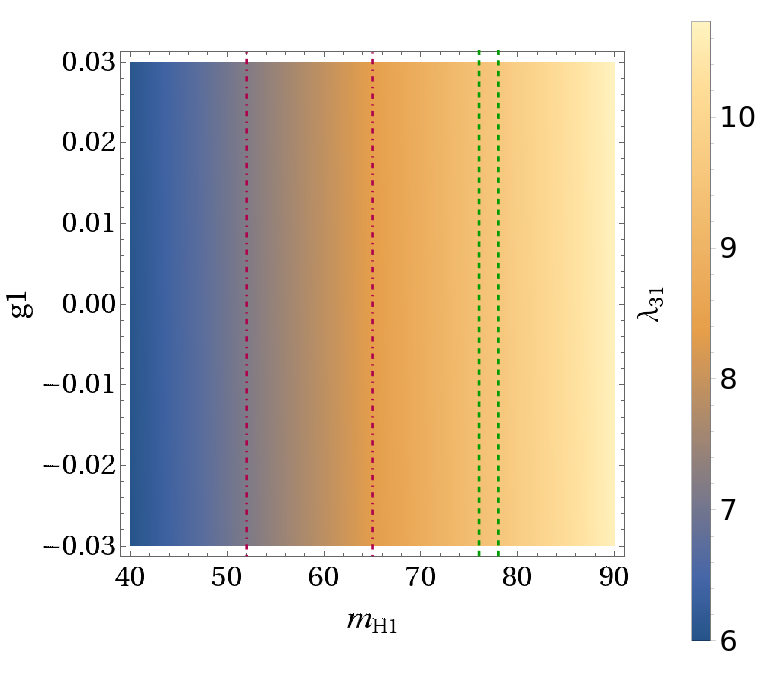}
		\includegraphics[scale=0.32315]{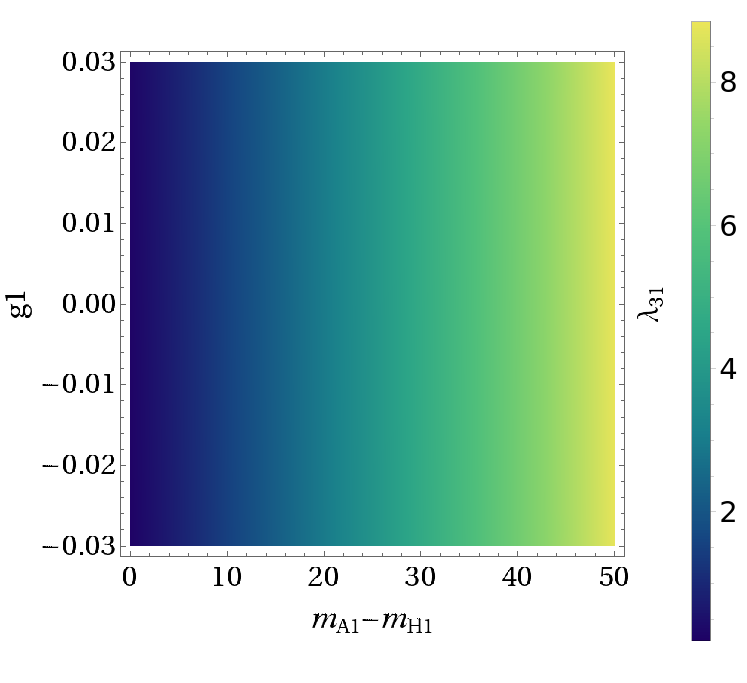}
		\caption{\label{scan1}Scan on the $\lambda^{(\prime)}$ parameters, for $ 0\leq\mu_{12}\leq63$ GeV. On the left, the scan is shown as a function of $g_1$ and $m_{H_1}$ with $ 40 $ GeV$\leq m_ {H_1} \leq 90 $ GeV, wherein the dotted lines show the intervals for $m_{H_1}$ that satisfy the experimental and theoretical restrictions. On the right, the scan is shown as a function of  $g_{1}$ and $\Delta_h(=m_{A_1}-m_{H_1})$ with $ 0 \leq m_{A_1}-m_{H_1}\leq 50 $ GeV.}
	\end{figure}
\end{center}

\begin{center}
	\begin{figure}[!t]
		\centering
		\includegraphics[scale=0.32315]{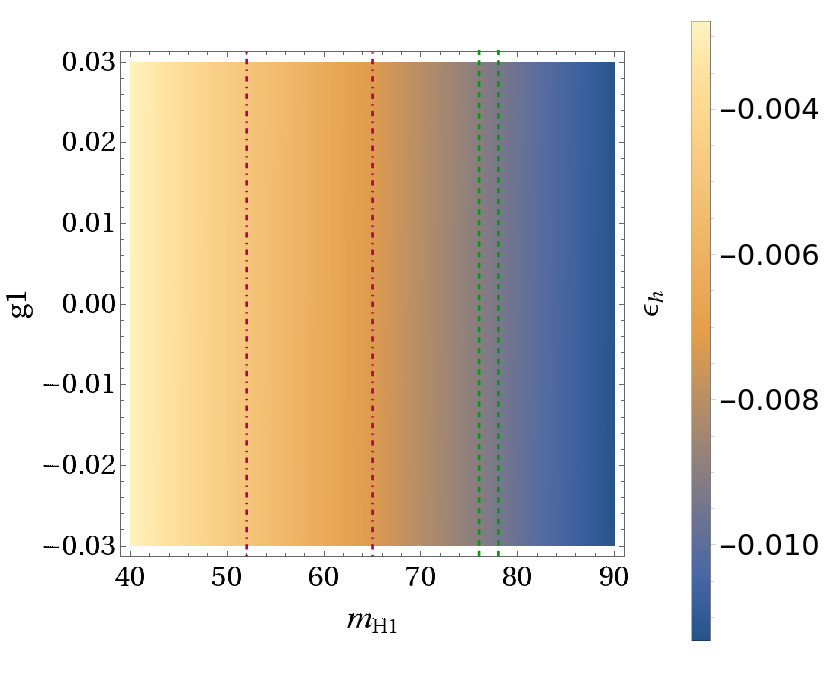}
		\includegraphics[scale=0.32315]{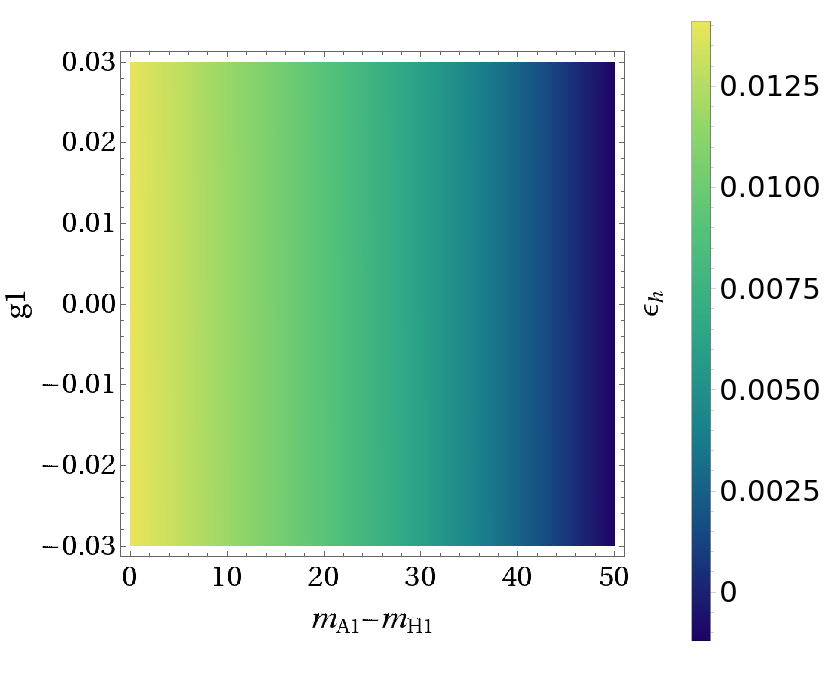}
		\includegraphics[scale=0.32315]{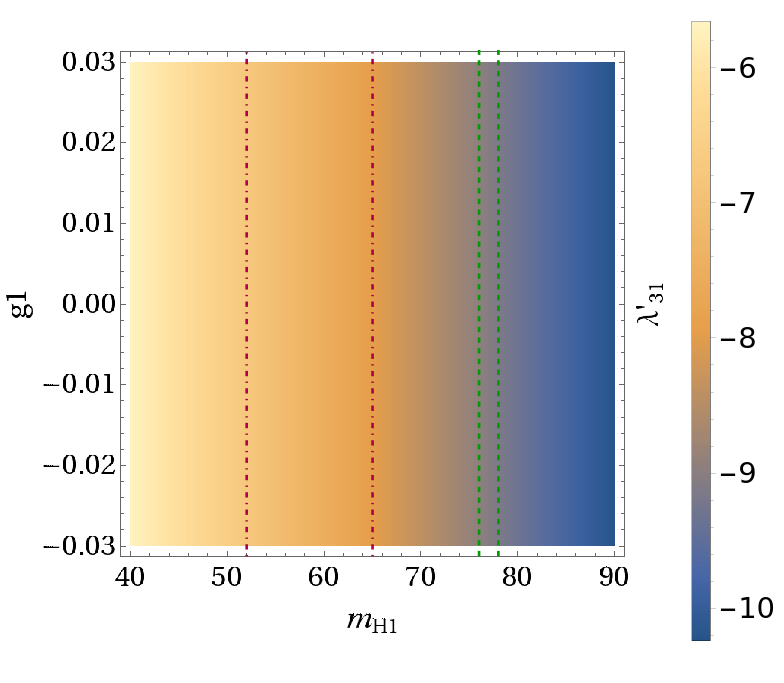}
		\includegraphics[scale=0.32315]{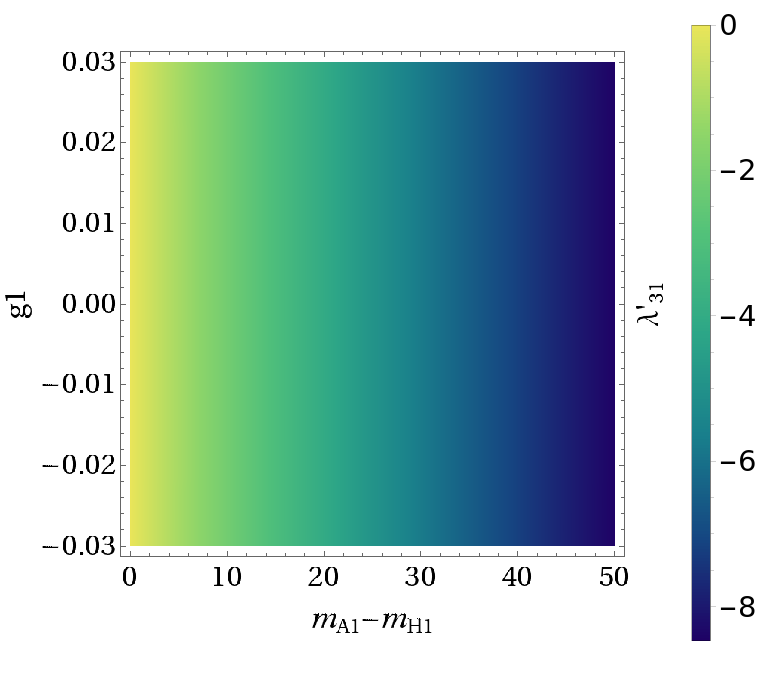}
		\includegraphics[scale=0.32315]{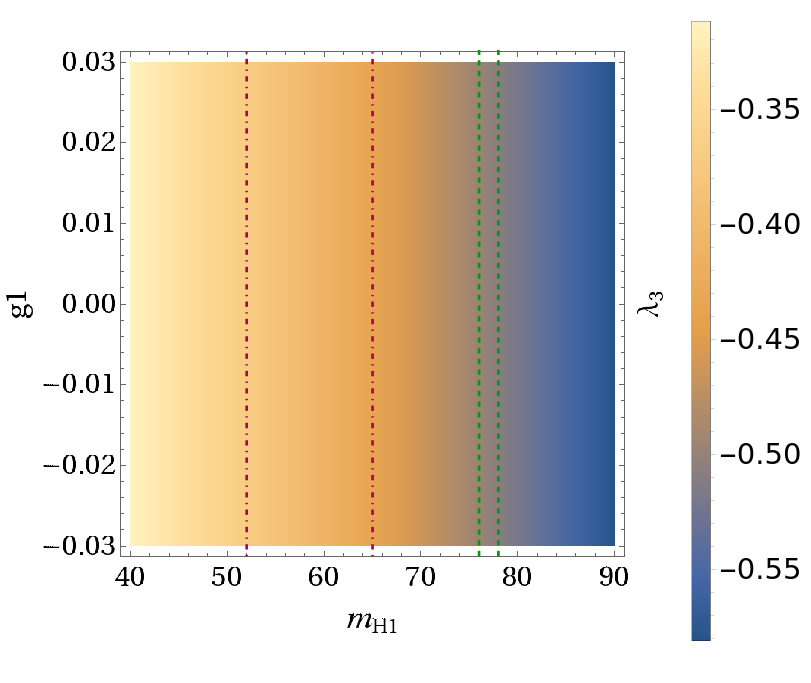}
		\includegraphics[scale=0.32315]{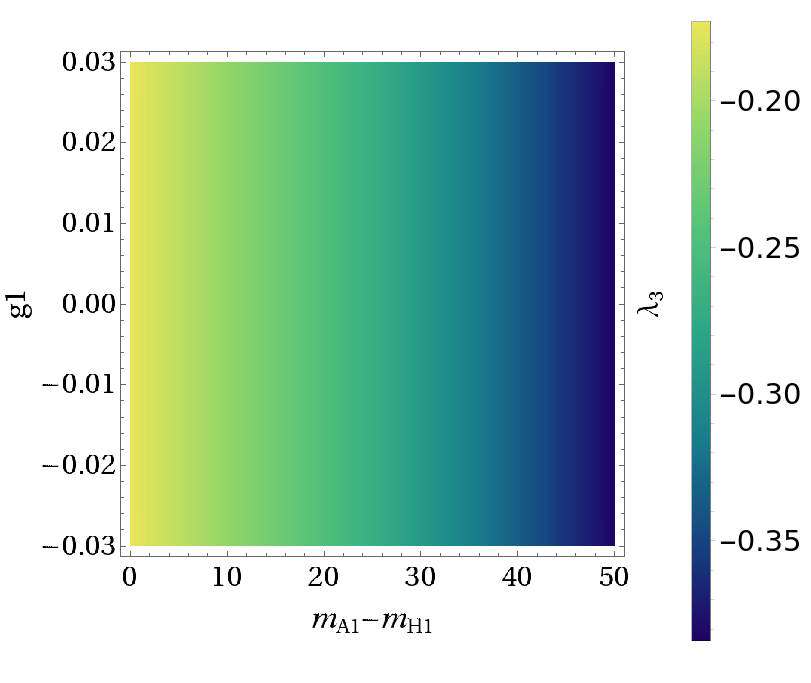}
		\caption{\label{scan2}Scan on the  $\lambda^{(\prime)}$ parameters and $\epsilon_n$,  for $ 0\leq\mu_{12}\leq63$ GeV. On the left, the scan  is shown as a function of $g_1$ and $m_{H_1}$ with $ 40 $ GeV$\leq m_ {H_1} \leq 90 $ GeV, wherein the dotted lines show the intervals for $m_{H_1}$ that satisfy experimental and theoretical restrictions. On the right, the scan is shown as a function of  $g_{1}$ and $\Delta_h(=m_{A_1}-m_{H_1})$  with $ 0 \leq m_{A_1}-m_{H_1}\leq 50 $ GeV.}
	\end{figure}
\end{center}

Assuming the parameter relations given in  Eqs. (\ref{m1})--(\ref{m3}) and (\ref{g12})--(\ref{par1}), we consider first the DM direct and indirect detection constraints as well as the theoretical ones. Figs. \ref{scan1} and  \ref{scan2} show a scan of the allowed values for the  $\lambda^{(')}$ (and $\varepsilon$) parameters for a fixed $\mu_{12}$ interval. The scanning is done over the range $40$ GeV $\leq m_{H_1} \leq 90$ GeV and mapped over two planes, $(m_{H_1}, g_1)$ and $(\Delta_h=m_{A_1}-m_{H_1}, g_1)$. (See Tab. \ref{Tab:parscan0} for the full list of  input values used for the relevant parameters.)

\begin{center}
	\begin{figure}
		\centering
		\includegraphics[scale=0.4]{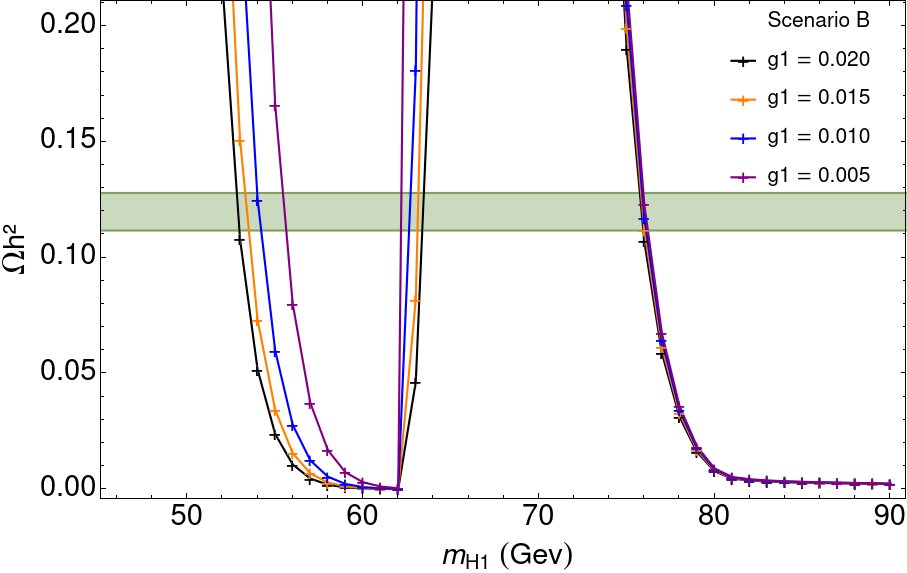}
		\caption{{\label{scanRD}\small The combined relic density of the DM constituents $H_1$ and $A_1$ with respect to $m_{H_1}$ in  scenario B given as lines for discrete  values of a positive $g_1$ coupling. The gray band represents the values for relic density that satisfy all constrains including direct and indirect detection as well as invisible decays. }}
	\end{figure}
\end{center}

\begin{center}
	\begin{figure}
		\centering
		\includegraphics[scale=0.4]{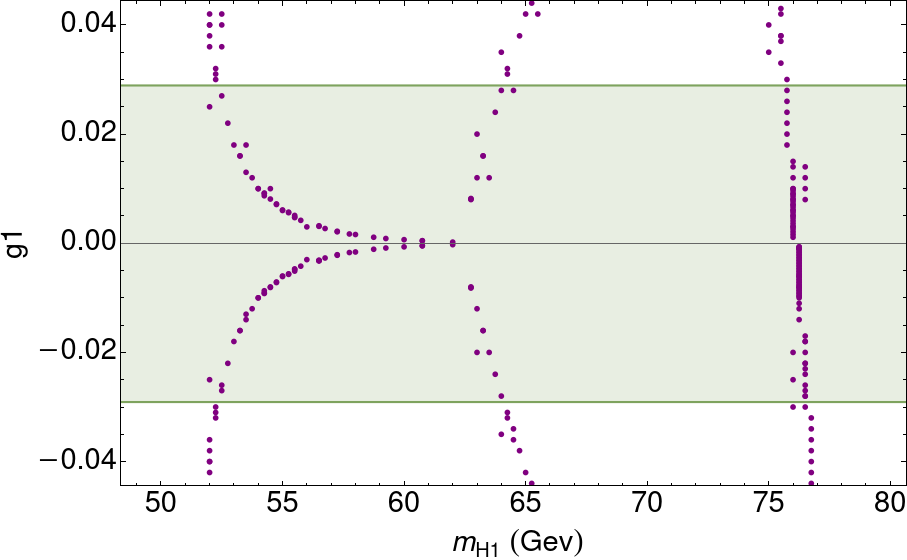}
		\label{mic1}
		\caption{{\small The combined relic density of the DM constituents  with respect to $m_{H_1}$ in  scenario B given as points for continuous values of the  $g_1$ coupling. The gray band represents the values for relic density that satisfy all constrains including direct and indirect detection as well as invisible decays.}}
	\end{figure}
\end{center}

For the numerical evaluation of the DM abundance, we have used micrOMEGAs \cite{Belanger2014}, which produced the results in
Fig. \ref{scanRD}, showing a scan of the combined relic density of the two components of DM, $ H_1 $ and $ A_1 $, for the  parameters shown in Tab. \ref{Tab:parscan}
(so-called scenario B). Different discrete values were used for $ g_1 $ here, with the $\Omega_{\rm DM} h^2$ predictions that fall within the green band representing the observed DM relic density within $3\sigma$. We can see that, in the mass ranges  $ 53 $ GeV $\leq m_{\rm H_1} \leq 64  $ GeV and $ 75 $ GeV$\leq m_{\rm H_1} \leq 78$ GeV, there is relic density saturation, which is clearly of interest. We further map the latter in Fig.~\ref{mic1}, wherein $g_1$ is allowed to vary over a continuous range (with the green band
again identifying the observed DM relic density within $3\sigma$). Altogether we notice that it is largely $g_1$ that dictates the model behaviour in relation to the relic abundance of DM.
Further analysis shows that, by varying $g_2$, the coefficient of the $hH_1H_2$ vertex, negative values for it affect the behaviour of the model in such a way that better results for the relic density are found, while the vertex coefficient $hH_1H_1$ is strongly constrained by invisible Higgs decays to have values $-0.029\leq g_1\leq 0.029$ (hence the choice of $y$-axis
in Figs. \ref{scan1} and  \ref{scan2}). 

\begin{table}[ht]
	\begin{center}
		\caption{\small Input  values for the parameter space scans in Figs. \ref{scan1}--\ref{scan2}.}
		\label{Tab:parscan0}
		\vspace{5mm}
		\begin{tabular}{|l|}
			\hline
			\multicolumn{1}{|c|}{Parameters}\\ \hline
			$40$ GeV$\leq m_{H_1}\leq 90$ GeV\\
			$\Delta_h=m_{A_1}-m_{H_1}=50$ GeV\\
			$\Delta_c=m_{H_1^{\pm}}-m_{H_1}=60$ GeV\\
			$\delta_c=m_{H_2^{\pm}}-m_{H_1^\pm}=10$ GeV\\
			$0.005\leq g_1 \leq 0.02$ and $g_2=-0.13$\\
			$0\leq \mu_{12}\leq 63$ GeV\\
			\hline
		\end{tabular}
	\end{center}
\end{table}

\begin{table}[ht]
	\caption{\small Input  values for the parameter space scan in Fig.~\ref{scanRD}.}
	\label{Tab:parscan}
	\begin{tabular}{|ccc|c|}
		\hline
		\multicolumn{3}{|c|}{Parameters}& scenario B \\
		\hline
		$\lambda_{11}=0.13$ & $\lambda'_{12}=0.12$ & $-0.029\leq g_1\leq 0.029$ & $m_{H_2}-m_{A_2} = $ 30 GeV 
		\\
		$\lambda_{22}=0.11$ & $\lambda_{1}=0.1$ &  $-0.2<g_2<0.2$ & $\Delta_c=$60 GeV 
		\\
		$\lambda_{12}=0.12$ & & $-0.1<\lambda_{2}<0.1$ &  $\delta_c=$10 GeV
		\\
		&&$\Delta_{n}=m_{A_2}-m_{H_1}$&$m_{A_1}-m_{H_1}=50$ GeV 
		\\
		&&$\Delta_{n'}=m_{H_2}-m_{A_1}$&$g_2=-0.13$
		\\
		&&$50$ GeV $<\Delta_{n}<100$ GeV &$\lambda_2=0.1$
		\\
		\hline
	\end{tabular}
\end{table} 

\section{Numerical results}
\begin{figure}[tbh]
	\centering
	\includegraphics[width=0.22\textwidth]{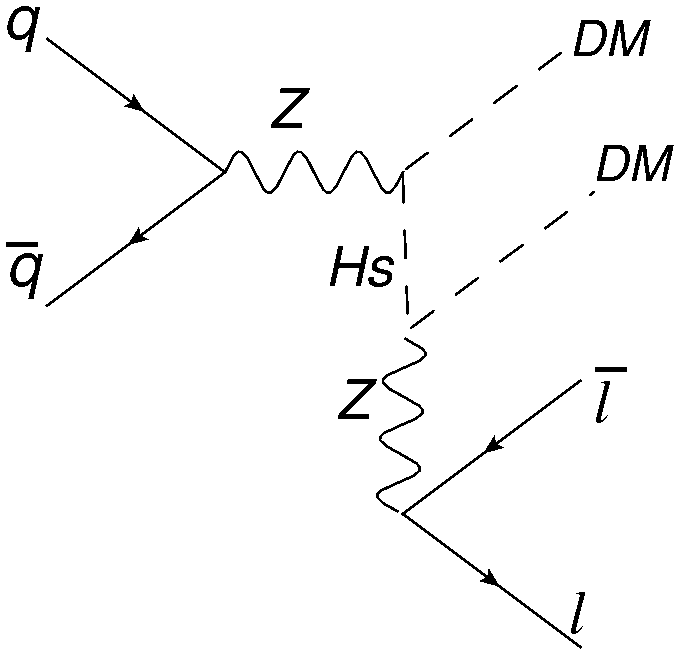} \hspace{.5cm}
	\includegraphics[width=0.21\textwidth]{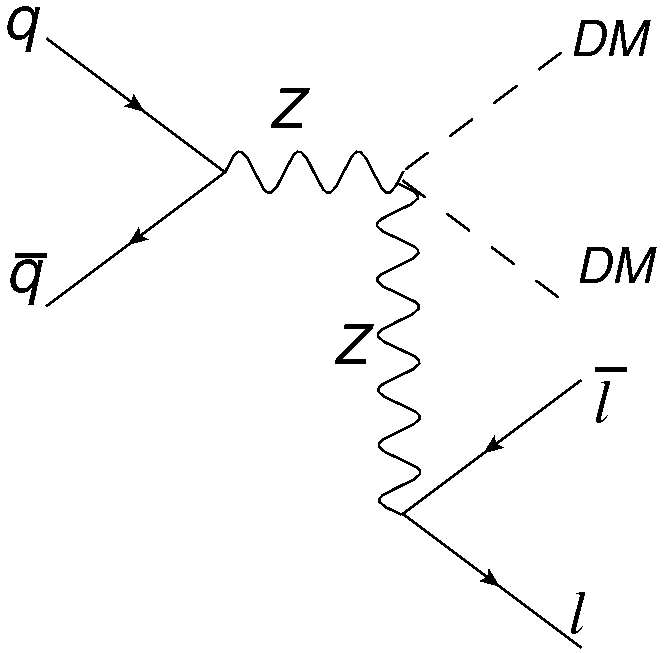}\hspace{.5cm}
	\includegraphics[width=0.3\textwidth]{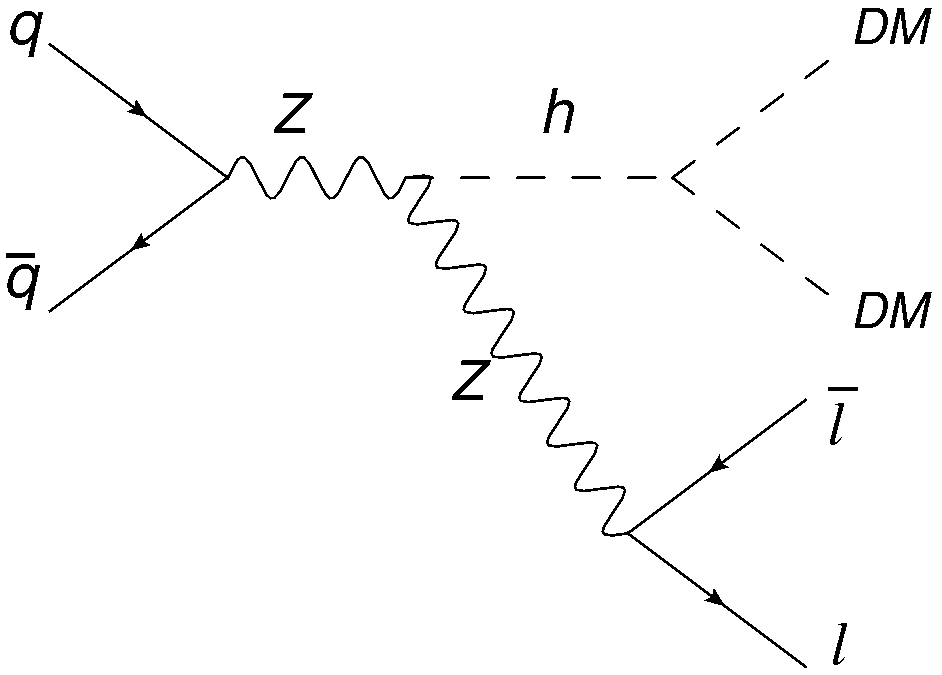} \hspace{.5cm}
	\caption{\label{qqto2l2h}Feynman diagrams for the processes $p p\to 2l+2{\rm DM}$, where $q=u,d$, $\bar{q}=\bar{u},\bar{d}$, $l = e ^ {-} (\mu^-) $ and $ \bar {l} = e ^ {+} (\mu^+) $, $H_s=A_2(H_2)$ and ${\rm DM}=H_1(A_1)$. The first diagram is the leading one.}
\end{figure}

\begin{figure}[tbh]
	\centering
	\includegraphics[width=0.22\textwidth]{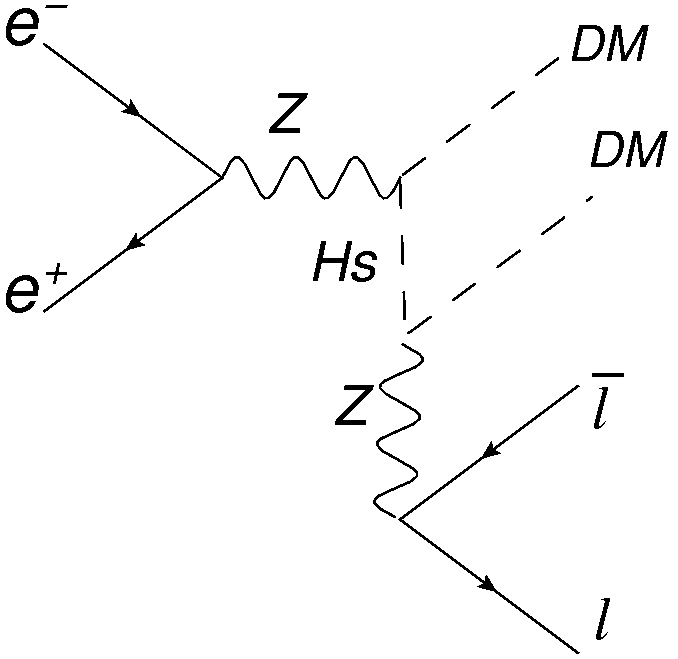} \hspace{.5cm}\vspace{.5cm}
	\includegraphics[width=0.21\textwidth]{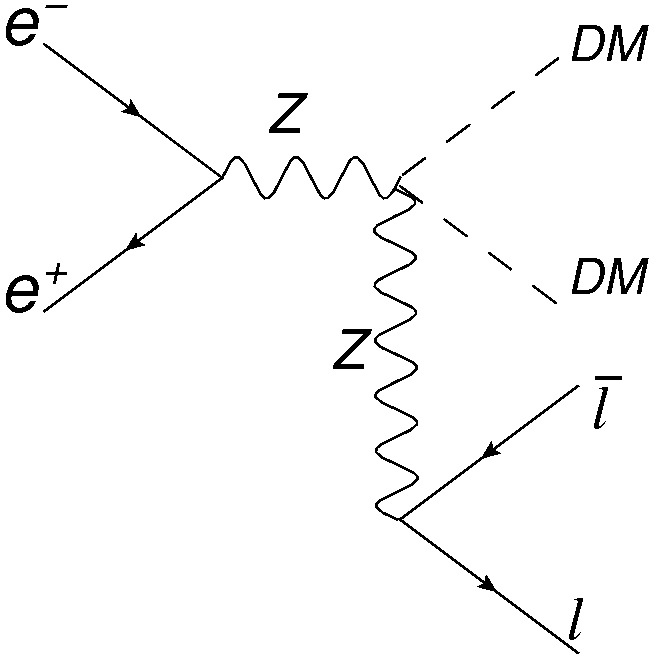}\hspace{.5cm}
	\includegraphics[width=0.3\textwidth]{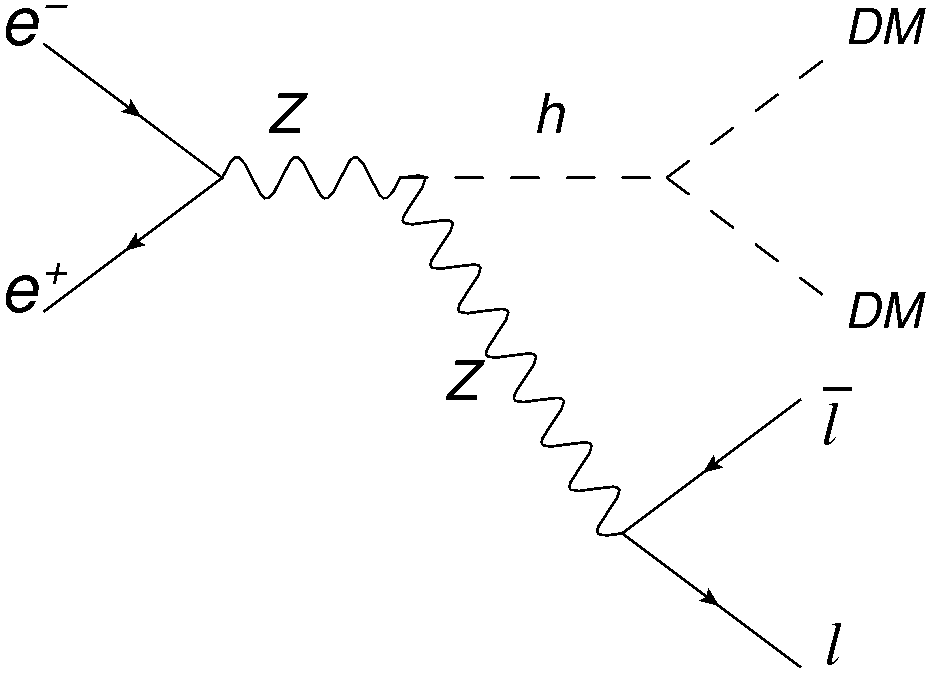} \hspace{1.5cm}
		\includegraphics[width=0.27\textwidth]{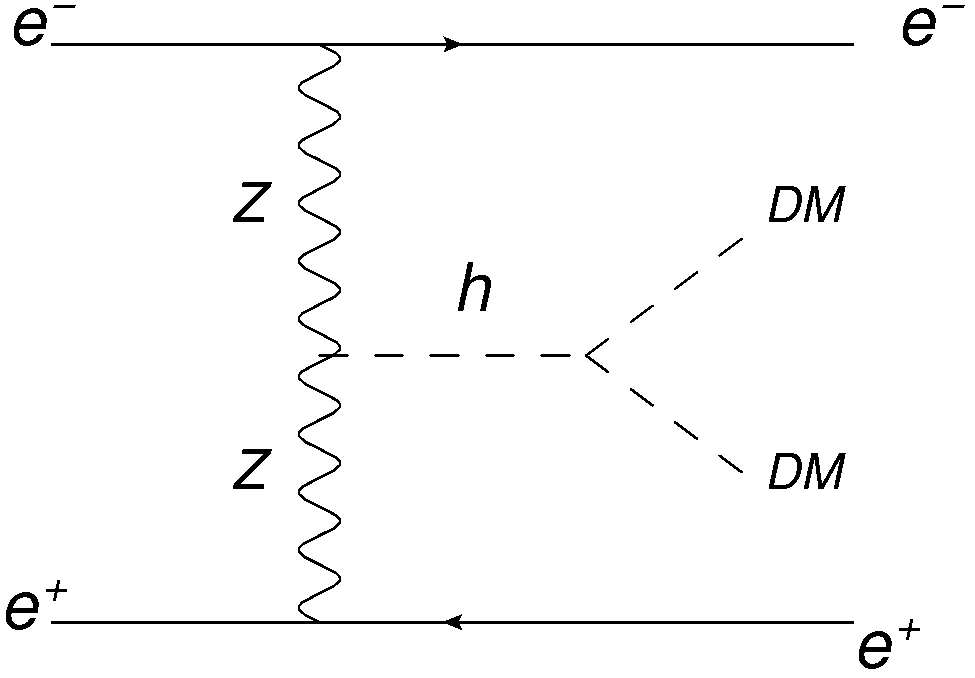} \hspace{.5cm}
	\includegraphics[width=0.21\textwidth]{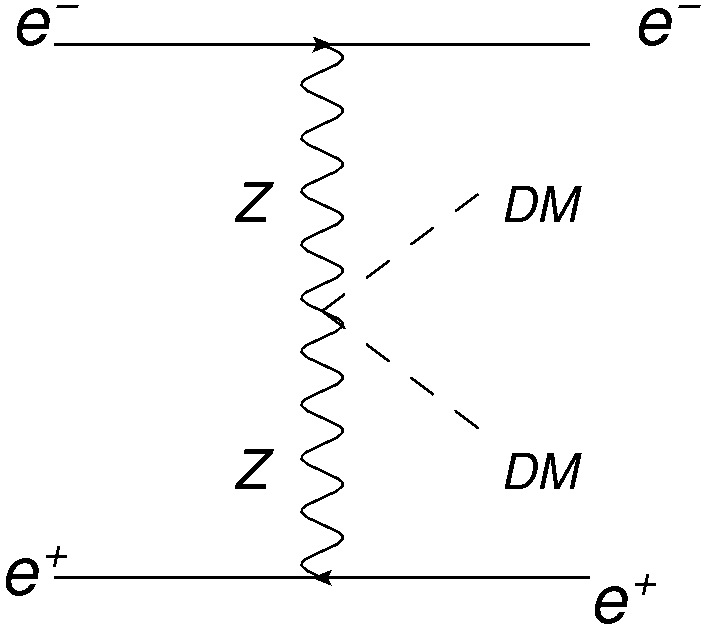}\hspace{.5cm} 
	\includegraphics[width=0.21\textwidth]{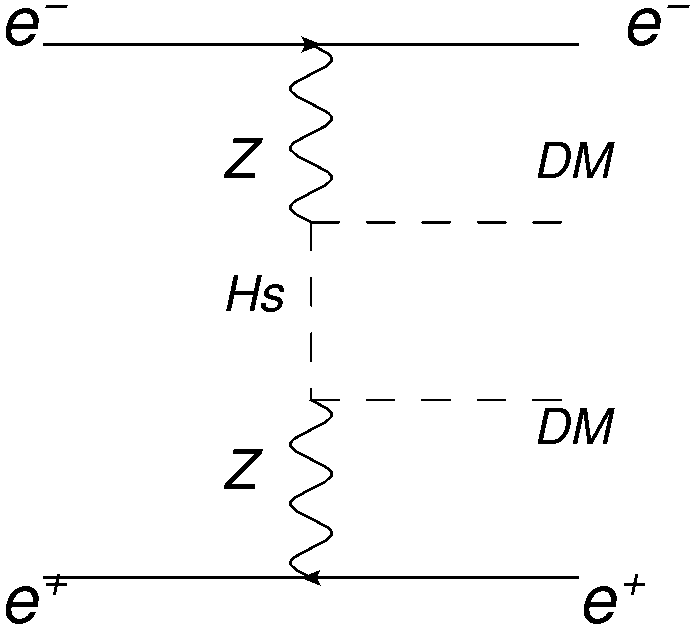} 
	\caption{\label{eeto2l2h}Feynman diagrams for the processes $e^+e^-\to 2l+2{\rm DM}$, where $l = e ^ {-} (\mu^-) $ and $ \bar {l} = e ^ {+} (\mu^+) $, $H_s=A_2(H_2)$ and ${\rm DM}=H_1(A_1)$. The first and last diagrams are the leading ones. (The diagrams in the second row only enter for the case $l=e^-$ and $\bar l=e^+$.)}
\end{figure}

Since the Hermaphrodite DM states $ H_1 $ and $ A_1 $  are protected from decaying into SM particles by the conservation of  $ Z_3 $ symmetry,  the two particles are stable in this model. According to Eqs. (\ref {m1})--(\ref {m2}) and (\ref {Dn}), $ m_ {H_1} <m_ {A_1} <m_ {A_2} <m_ {H_2} $. The heavier particles, $ A_2 $ and $ H_2 $, are unstable and their decays in $ H_1 $, $ A_1 $ or SM particles could provide experimental signals. 
We focus our study on the production of DM, considering the following processes:  $ p p \to 2l + 2 {\rm DM}$ for the current LHC machine (see the Feynman diagrams in  Fig. \ref{qqto2l2h}) and  $ e ^ + e^- \to 2l +  2 {\rm DM}$ for a future electron-positron accelerator (see the Feynman diagrams in  Fig. \ref{eeto2l2h}), wherein $ l (\bar {l}) = e ^ + (e ^ -), \mu ^ + (\mu ^ -) $ and $ {\rm DM} = H_1, A_1 $.   (Hereafter, we use the notation $2l$ for short, to signify an electron or muon pair of opposite charge.)  The leading diagrams are those where the vertices $ Z  A_1H_2 $ and $ Z A_2H_1 $ appear.  The aforementioned processes   to obtain DM particles also involve decays of  the $Z$ boson and the $h$ state. As mentioned, we take the scalar field $h$ to the SM-like Higgs boson, hence $m_h =125$ GeV. As for the $H_i,A_i$ dark states, we will select a few BPs from
the available parameter space that we have isolated. Finally,  MadGraph will be used for our calculations at the parton level \cite{Alwall:2014hca} with integrated and differential distributions obtained via MadAnalysis  \cite{Conte:2012fm}.
\subsection{LHC signatures  $p p \to 2 l + 2 H_1 (2 A_1)$}
For the LHC machine, we consider the following signature:  $  2 l +  \cancel{E}_T$ (where $\cancel{E}_T$ is the 
missing $E_T$  of the event),  which can be  induced at tree-level by the processes  
$p p \to 2 l + 2 H_1 (2 A_1)$.
In our analysis, the following basics cuts on leptons are considered: transverse momentum $ p_T(l) >10$ GeV,
pseudorapidity  $|\eta(l)| < 2.5 $ and separation $\Delta R(l^+,l^-)>0.4$.
We take one BP within  scenario B (with $\Delta_{n(n')}=70(50)$ GeV), which 
is identified  by the following mass and  coupling parameter values:  $m_{H_1} = 53 $ GeV, $m_{A_1} = 103 $ GeV, 
$m_{A_2} = 123 $ GeV, $m_{H_2} = 153 $ GeV and $g_1= 0.029$. Since the mass difference $\Delta_h=m_{A_1} - m_{H_1}$  is taken  
larger than the detector resolution in $\cancel{E}_T$ as well as in mass variables (e.g.,   invariant, $M(l^+l^-)$,  or  transverse, $M_T(l^+l^-)$) involving leptons (and $\cancel{E}_T$)
at the  LHC, we could then observe the presence of  both Hermaphrodite DM components simultaneously.  For this BP of scenario B, we show in  Tab. \ref{sigmah1a1} both cross-sections and  event rates at partonic level, assuming $\sqrt s_{pp}=14$ TeV and a luminosity of $L=100$ fb$^{-1}$, which could be accrued within a few
years of operation at the LHC during Run 3.   In general, the value of the cross-section $\sigma (p p \to 2 l + 2 H_1 (2 A_1) )$ depends mostly  on the mass difference  $\Delta_n=m_{A_2}-m_{H_1}$\;($\Delta_{n'}=m_{H_2}-m_{A_1}$), which is thus very important in the description of the  distributions generated.
\begin{table}[htp]
\caption{Cross-section  for the processes $p p \to 2 l + 2 {\rm DM}$, with ${\rm DM} = H_1, A_1$, taking $L=100$ fb$^{-1}$ as well as $m_{H_1} = 53 $ GeV, $m_{A_1} = 103 $ GeV,  $m_{A_2} = 123 $ GeV, $m_{H_2} = 153 $ GeV, $g_1=0.029$ with other parameters as in scenario B.}
\label{sigmah1a1}
\begin{center}
\begin{tabular}{|c|c|c|c|}
\hline
DM & $\sigma (p p \to 2 l + 2 {\rm DM}) $ & Event rates \\ \hline
$H_1 $ & 0.280 pb & $2.8 \times 10^{4}$ \\ \hline
$A_1 $ & 0.135 pb & $1.35 \times 10^{4}$ \\ \hline
\end{tabular}
\end{center}
\end{table}
\begin{center}
	\begin{figure}
		\centering
		\includegraphics[scale=0.5]{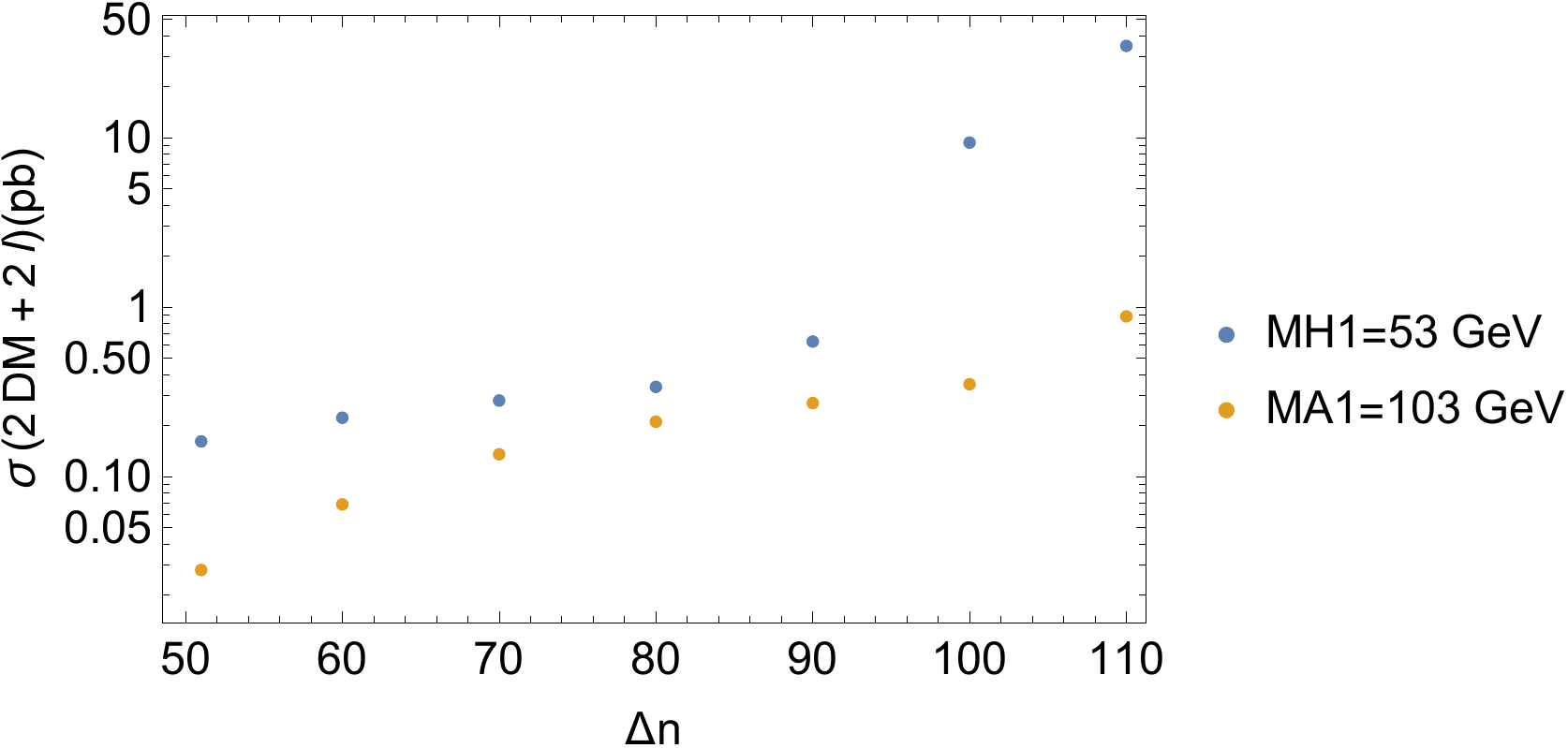}
		\caption{Cross-section of the processes $p p \to 2 l +2 H_{1}$ (blue colour) and $p p\to 2 l +2 A_{1}$  (orange colour) as a function of 
$\Delta_{n}=m_{A_2}-m_{H_1}$ (in GeV), with $ m_ {A_1} =103$ GeV and $ m_{H_1}= 53 $ GeV.  All points are compliant with DM (in)direct detection bounds  and relic density as well as the Higgs to invisible BR constraint. Here, $\sqrt s_{pp}=14$ TeV. }
		\label{Dnpp}
	\end{figure}
\end{center}
In Fig. \ref{Dnpp}, we show the behaviour of the cross-section $\sigma (p p \to 2 l + 2 H_1 (2 A_1) )$ versus the $\Delta_n=m_{A_2}-m_{H_1}$ parameter, the points in blue colour being for  $H_1 $  and the points in orange colour being for $A_1$. One can see that, in the region 70 GeV $\leq \Delta_n \leq 90$ GeV, the cross-sections for $H_1 $  and $A_1$ are closest to each other. A small relative rate, alongside a sufficiently large absolute value of either $\sigma$, is 
a precondition to observe the two Hermaphrodite candidates simultaneously. Hence, our BP   is taken in this range,  in 
 particular, with $\Delta_n = 70$ GeV, so that the events rates for the two DM processes are within the same order of magnitude (see Tab. \ref{sigmah1a1}).
\begin{figure}
  \begin{center}
    \includegraphics[scale=0.25]{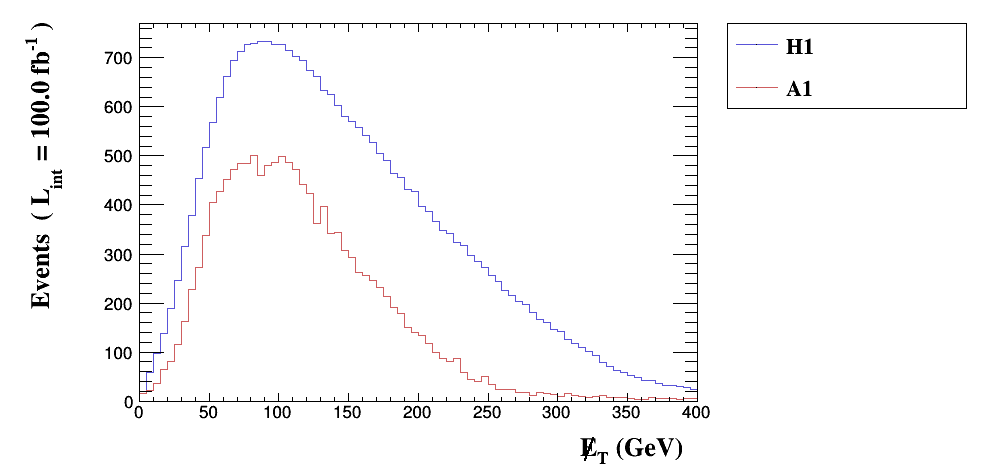}
     \includegraphics[scale=0.25]{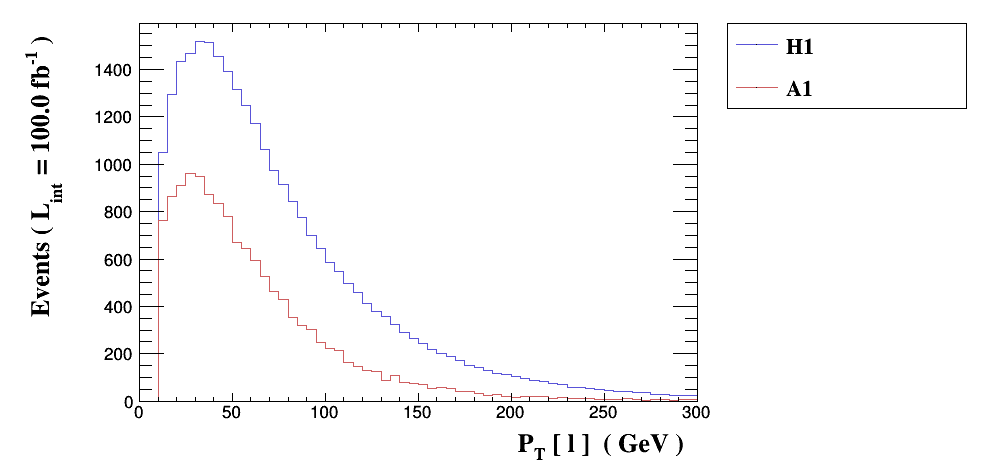}
\caption{Spectra in missing transverse energy (left) and transverse momentum of each lepton (right) for the processes $p p \to 2 l + 2 {\rm DM}$,  where the distributions are identified by a blue colour line for  ${\rm DM}=H_1$  and a red colour line for ${\rm DM}=A_1$. These correspond to scenario B with 
the specific choices  $\Delta_n =m_{A_2}-m_{H_1} =70$ GeV and  $\Delta_{n'}=m_{H_2}-m_{A_1} =50$ GeV.
} 
 \label{met-pp}
  \end{center}
\end{figure}
Following such a choice, in
  Fig. \ref{met-pp}, we obtain rather similar shapes in the distributions of the  missing transverse energy   $ \cancel{E}_T$ and  transverse momentum of each lepton  $p_T (l)$. 
\begin{figure}
  \begin{center}
    \includegraphics[scale=0.25]{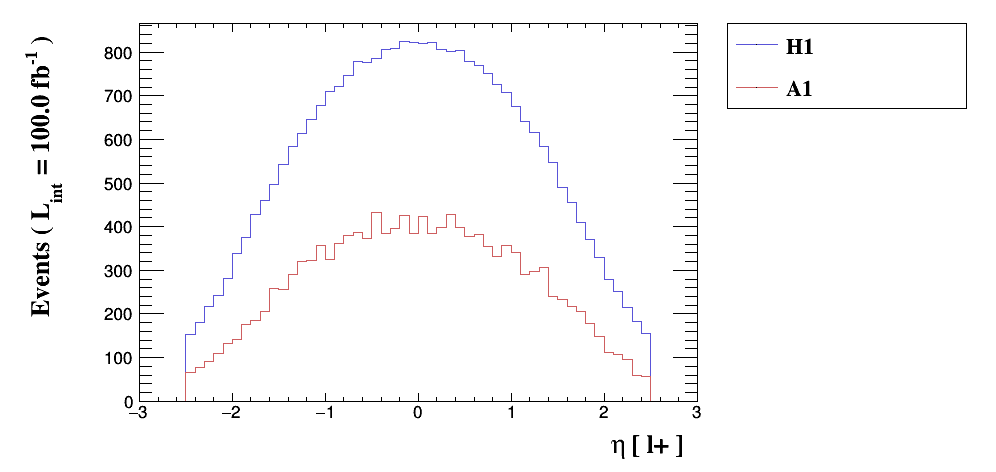}
      \includegraphics[scale=0.25]{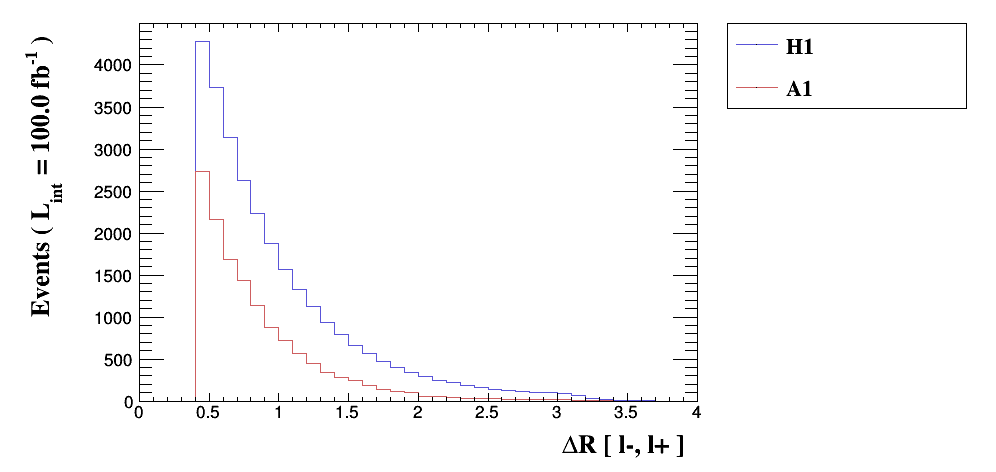}
 \caption{Spectra in pseudorapidity of each lepton (left) and relative distance between leptons (right)  for the processes $p p \to 2 l + 2 {\rm DM}$,  where the distributions are identified by  a  blue colour line for  ${\rm DM}=H_1$  and a red colour line for ${\rm DM}=A_1$. These correspond to scenario B with 
the specific choices  $\Delta_n =m_{A_2}-m_{H_1} =70$ GeV and  $\Delta_{n'}=m_{H_2}-m_{A_1} =50$ GeV.
}
  \label{eta}
  \end{center}
\end{figure}
\begin{figure}
  \begin{center}
    \includegraphics[scale=0.3]{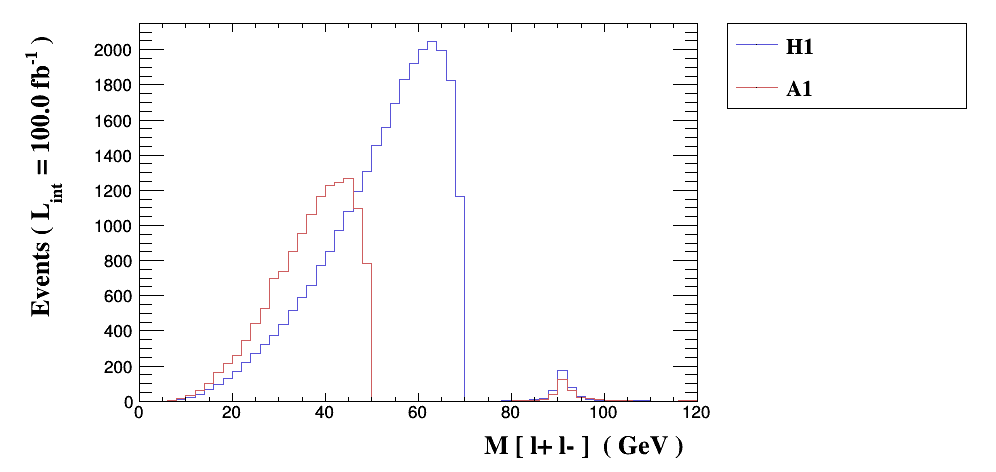}
\caption{Spectra in  invariant mass of the leptons for the processes $p p \to 2 l + 2 {\rm DM}$, where the distributions are identified by  a  blue colour line for  ${\rm DM}=H_1$  and a red colour line for ${\rm DM}=A_1$. These correspond to scenario B with 
the specific choices  $\Delta_n =m_{A_2}-m_{H_1} =70$ GeV and  $\Delta_{n'}=m_{H_2}-m_{A_1} =50$ GeV.
}
 \label{mll}
  \end{center}
\end{figure}
\begin{figure}
\begin{center}
    \includegraphics[scale=0.3]{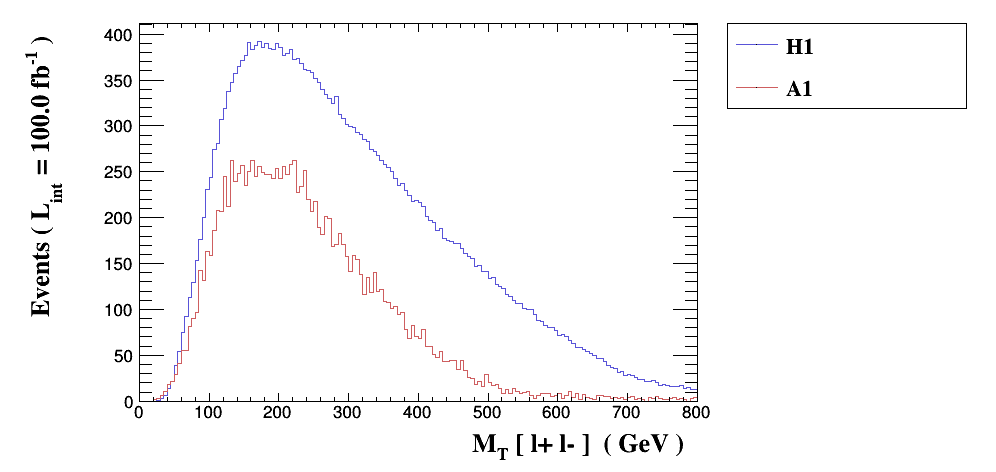}\\
\caption{Spectra in transverse mass of the final state for the processes $p p \to 2 l + 2 {\rm DM}$, where the distributions are identified by  a  blue colour line for  ${\rm DM}=H_1$  and a red colour line for ${\rm DM}=A_1$. These correspond to scenario B with 
the specific choices  $\Delta_n =m_{A_2}-m_{H_1} =70$ GeV and  $\Delta_{n'}=m_{H_2}-m_{A_1} =50$ GeV.
}
  \label{mT-lhc}
  \end{center}
\end{figure}
Likewise, in Fig.  \ref{eta},  one can see that also the pseudorapidity of each lepton $\eta(l)$   and separation between them $\Delta R(l^+l^-)$ have rather
similar shapes. As the last three variables are used for selection purposes, we would conclude that our envisaged cutflow would not dramatically change the
relative rates seen in Tab.~\ref{sigmah1a1}. Unfortunately, though, the $\cancel{E}_T$ spectrum cannot afford one with separating the $H_1$ and $A_1$ components of 
Hermaphrodite DM. However, this would become possible for the case of the invariant mass of the di-lepton system, as seen in  Fig. \ref{mll}, wherein the off-shell $Z$ large peaks for $H_1$ and $A_1$ are separately visible and can strongly be correlated to the values of $\Delta_n$ and $\Delta_{n'}$, respectively. (Also, notice the small on-shell $Z$ 
peaks for both DM candidates.)   In contrast, the transverse mass distribution $M_T^2(l^+ l^-) = (\sum_{i}^{l^+,l^-}  E_{Ti})^2 - (\sum_i^{l^+,l^-} p_{Ti})^2$, wherein the proton beams are along the $z$-axis, shown in  Fig.  \ref{mT-lhc} offers one little chance to pinpoint the presence of Hermaphrodite DM.
\subsection{Electron-positron collider processes  $e^+ e^- \to 2 l + 2 H_1 (2 A_1)$}
\begin{center}
	\begin{figure}
		\centering
		\includegraphics[scale=0.5]{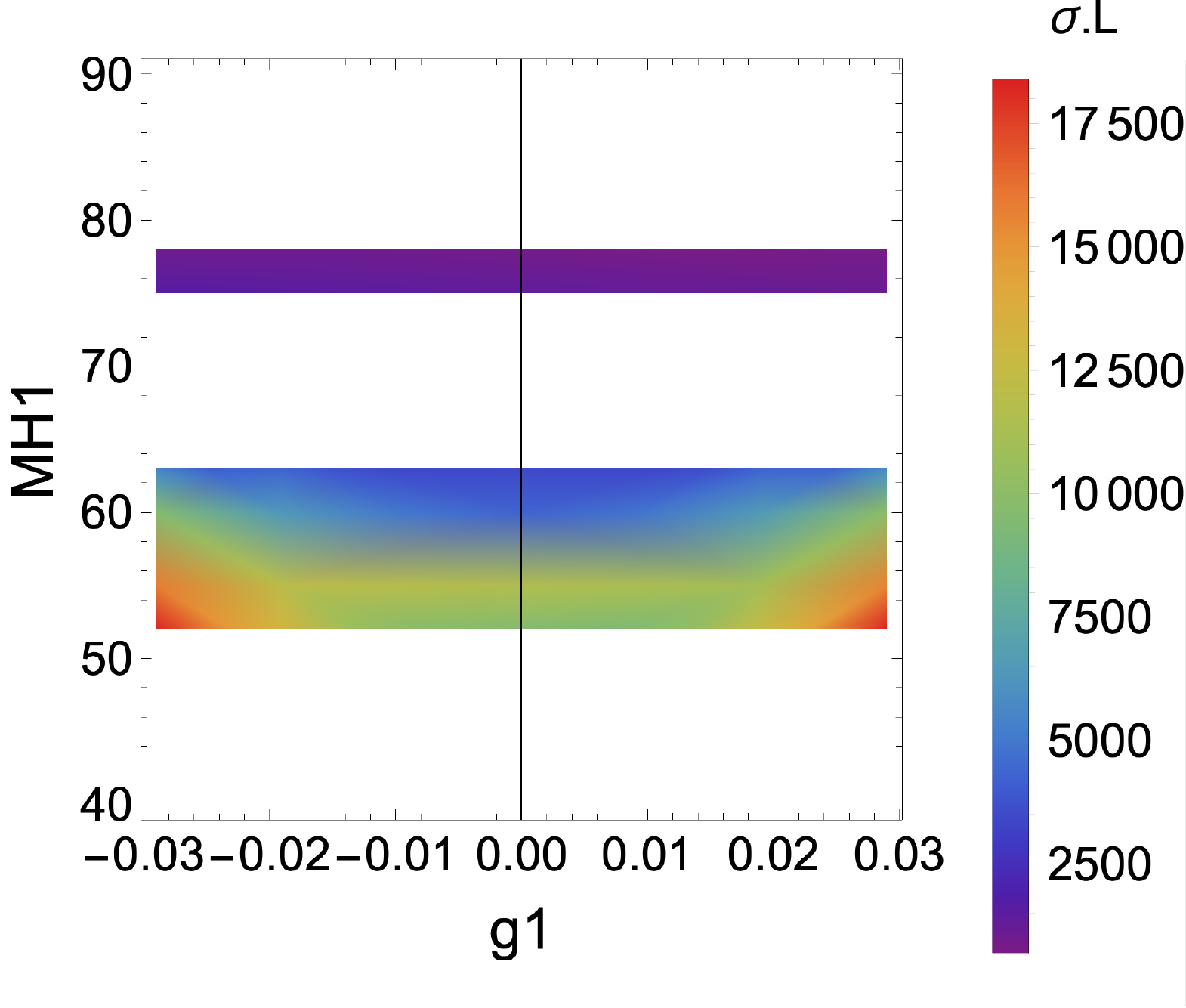}
		\caption{Event rates  of the process $e^+e^- \to 2 l +2 H_{1}$ as a function of $g_1$ and $M_{H_1}$ for $\sqrt s_{ee}=250$ GeV
and $L=1000$ fb$^{-1}$ over the parameter space corresponding to scenario B in Tab.~\ref{Tab:parscan}.  All points are compliant with DM (in)direct detection bounds  and relic density as well as the Higgs to invisible BR constraint. (The most optimistic BP is for $m_{H_1}=53 $ GeV and $g_1=-0.029$, where the cross-section is $\sigma = 16.01$ fb.) }
		\label{N1}
	\end{figure}
\end{center}

\begin{center}
	\begin{figure*}[tbh]
		\centering
		\includegraphics[scale=0.5]{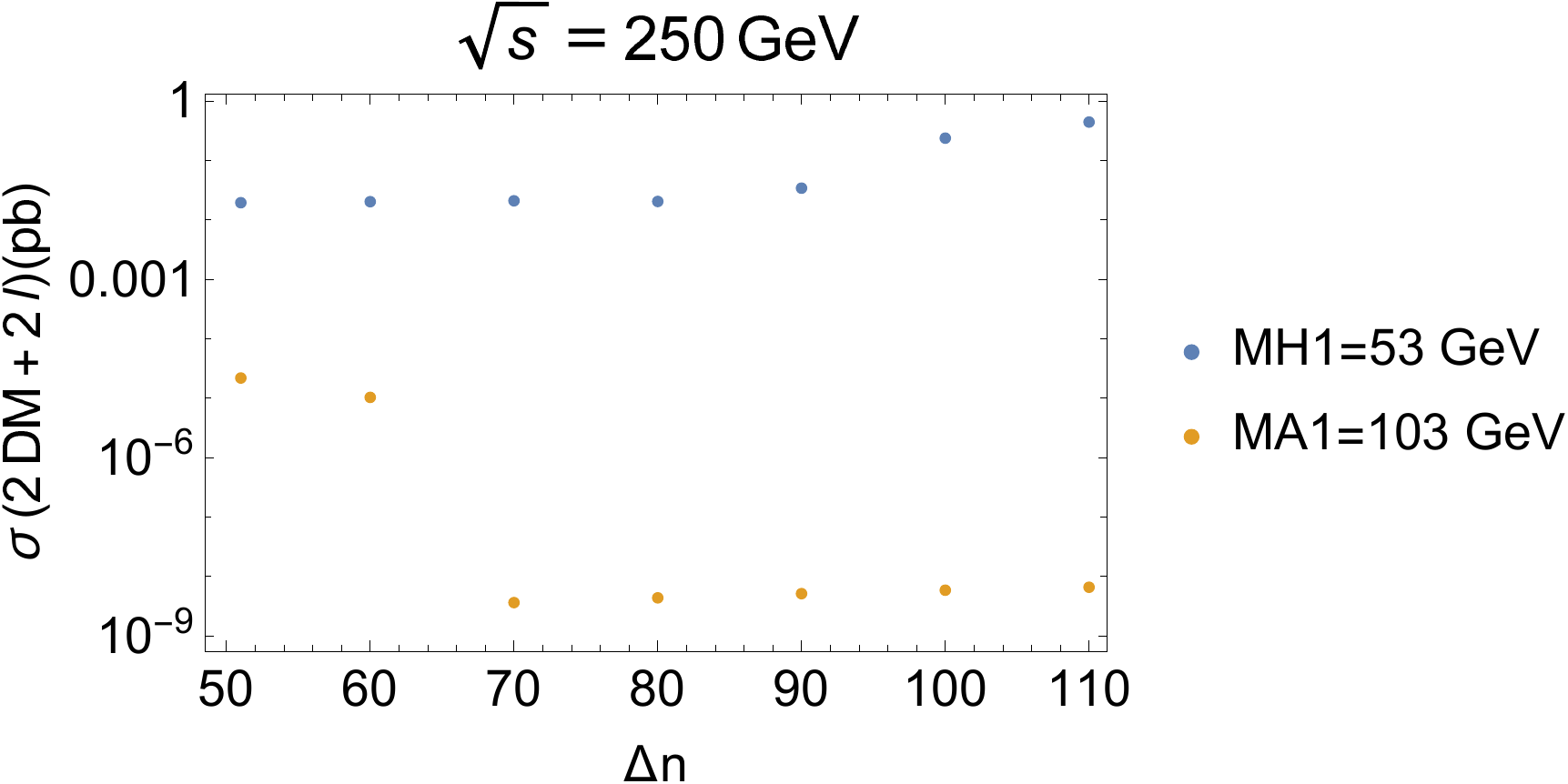}
		\includegraphics[scale=0.5]{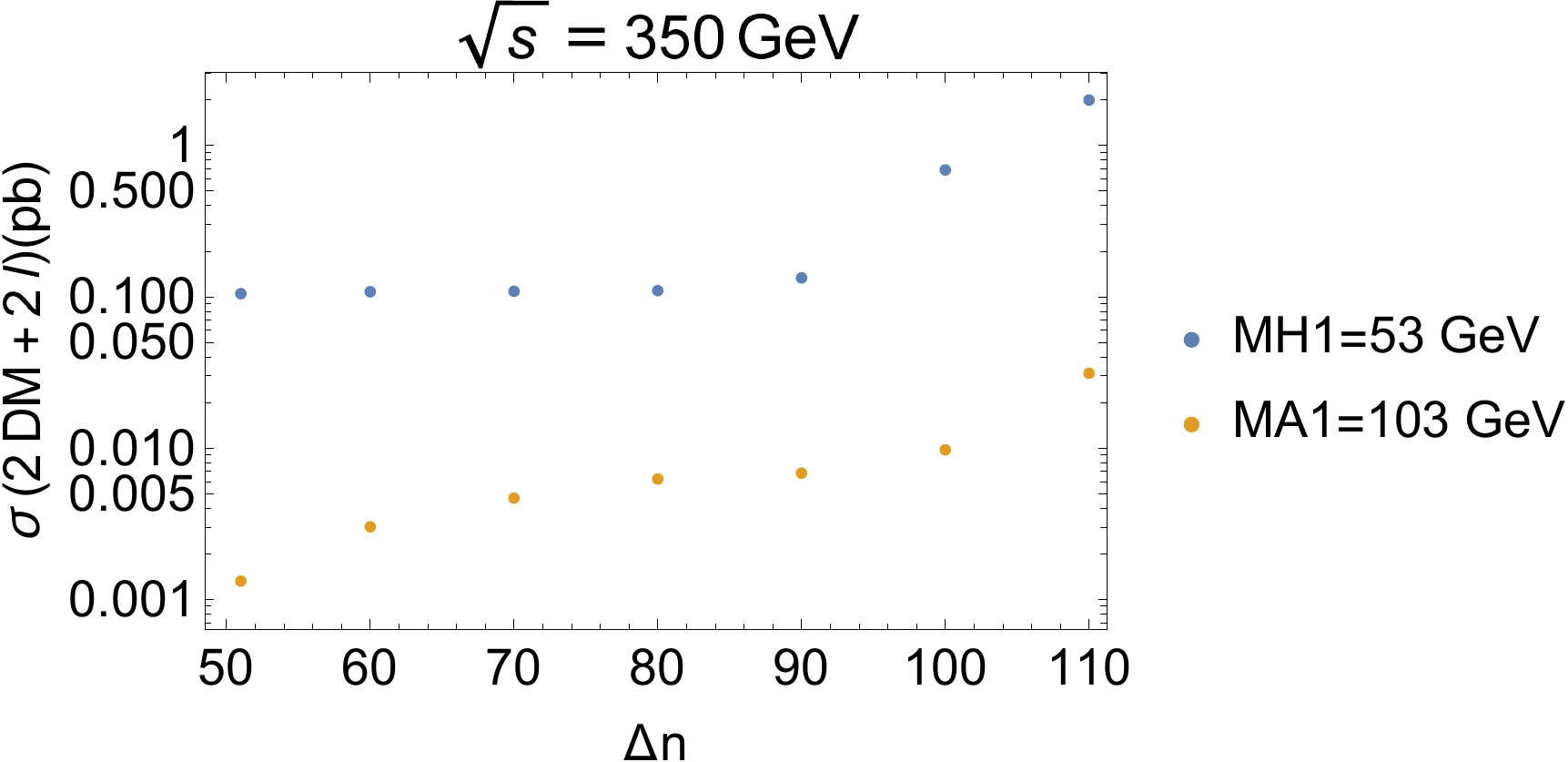}
		\includegraphics[scale=0.5]{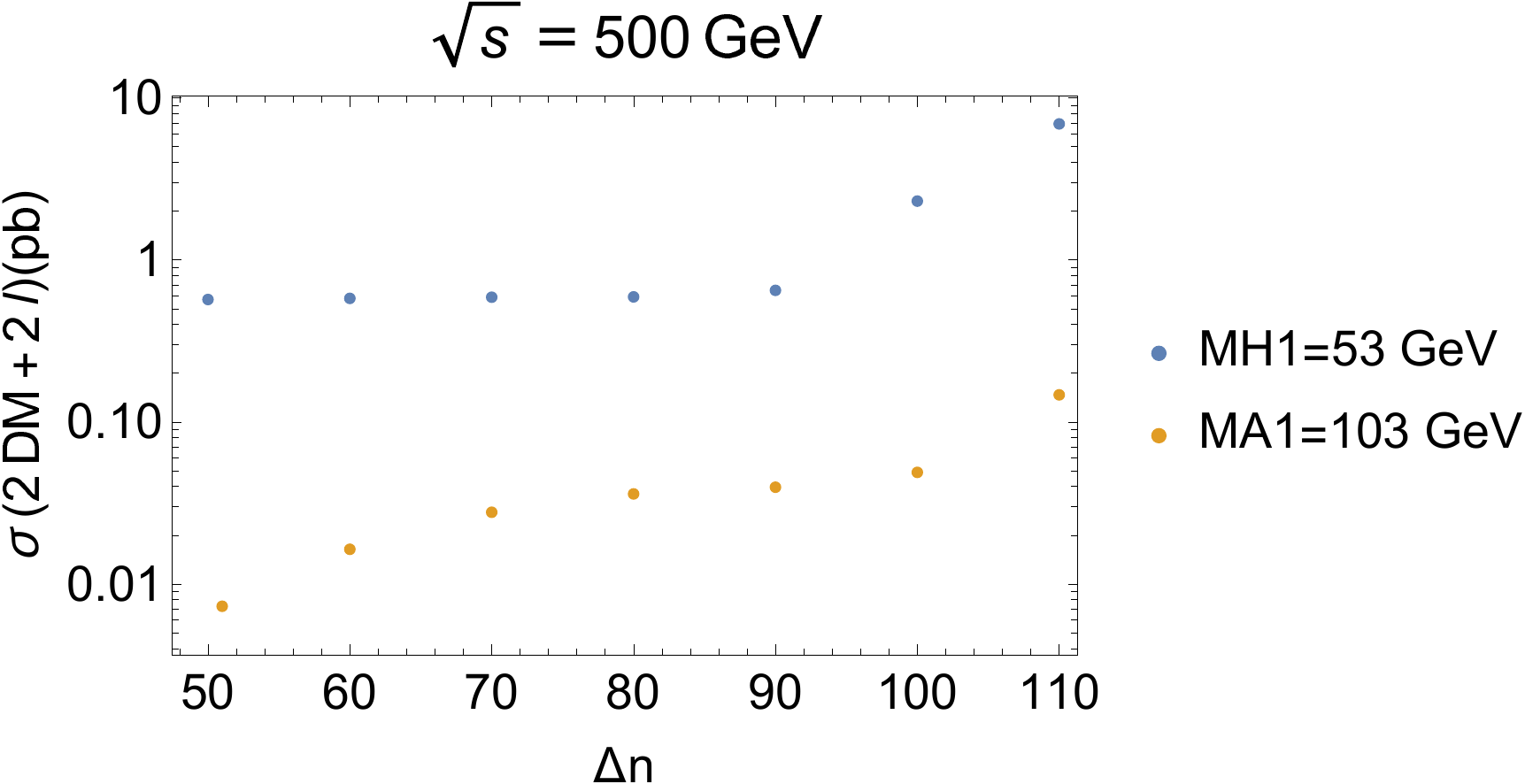}
		\includegraphics[scale=0.5]{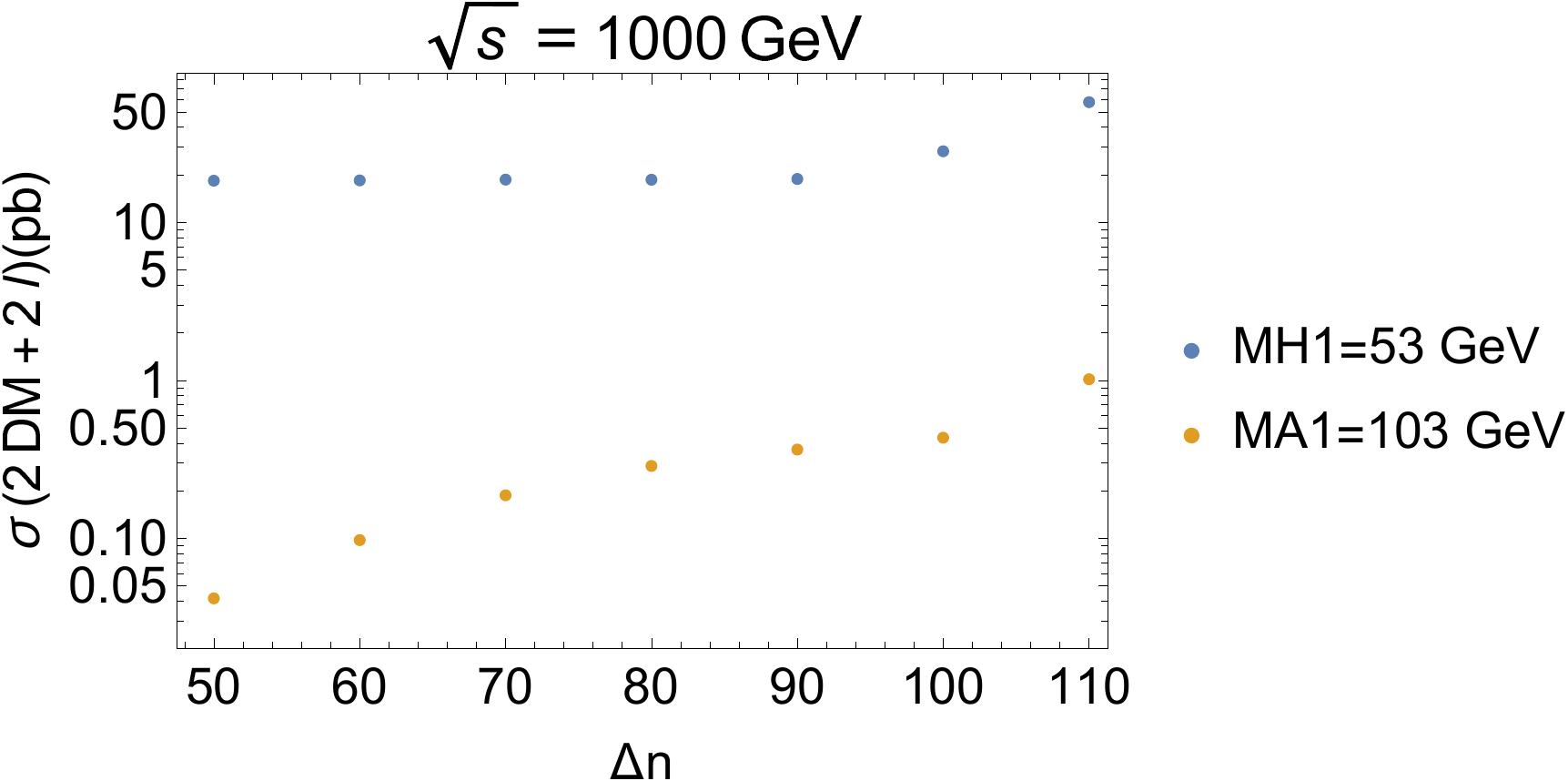}
		\caption{Cross-section of the processes $e^+e^- \to 2 l +2 H_{1}$ (blue colour) and $e^+e^-\to 2 l +2 A_{1}$  (orange colour) as a function of 
$\Delta_{n}=m_{A_2}-m_{H_1}$ (in GeV), with $ m_ {A_1} =103$ GeV and $ m_{H_1}= 53 $ GeV.  All points are compliant with DM (in)direct detection bounds  and relic density as well as the Higgs to invisible BR constraint. Here, $\sqrt s_{ee}=250$ (top-left), 350 (top-right), 500 (bottom-left) and 1000 (bottom-right) GeV.}
		\label{Deltan-ILC}
	\end{figure*}
\end{center}

\begin{figure}
  \begin{center}
    \includegraphics[scale=0.25]{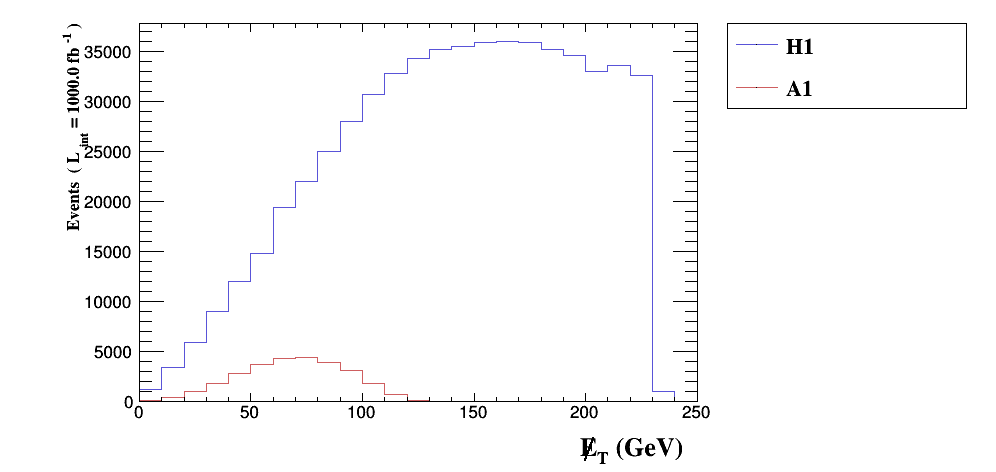}
     \includegraphics[scale=0.25]{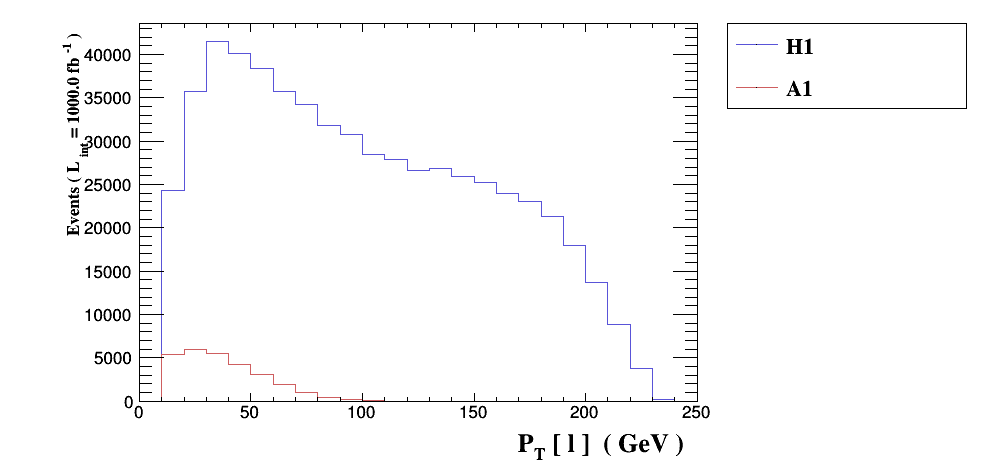}
\caption{Spectra in missing transverse energy (left) and transverse momentum of each lepton (right) for the processes $e^+e^-\to 2 l + 2 {\rm DM}$,  where the distributions are identified by a blue colour line for  ${\rm DM}=H_1$  and a red colour line for ${\rm DM}=A_1$. These correspond to scenario B with 
the specific choices  $\Delta_n =m_{A_2}-m_{H_1} =70$ GeV and  $\Delta_{n'}=m_{H_2}-m_{A_1} =50$ GeV.
}
 \label{met-ILC-1}
  \end{center}
\end{figure}

\begin{figure}
  \begin{center}
    \includegraphics[scale=0.25]{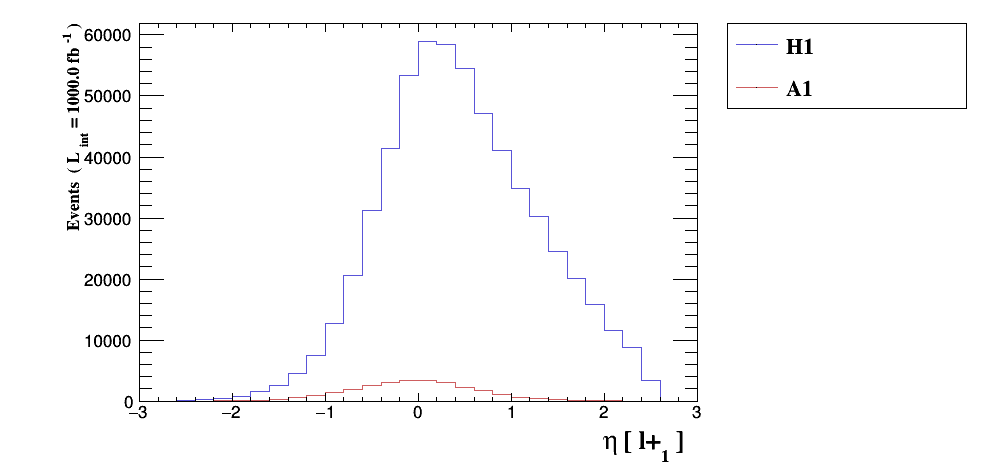}
     \includegraphics[scale=0.25]{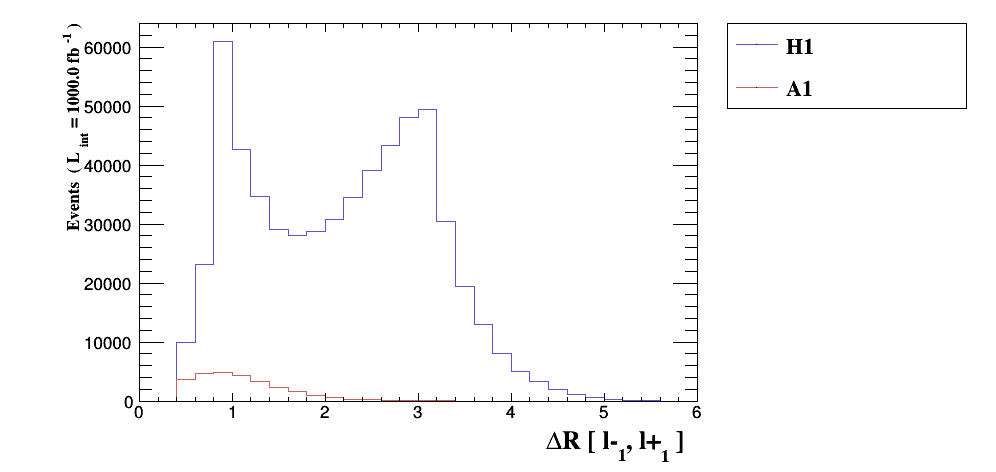}
 \caption{Spectra in pseudorapidity of each lepton (left) and relative distance between leptons (right)  for the processes $p p \to 2 l + 2 {\rm DM}$,  where the distributions are identified by  a  blue colour line for  ${\rm DM}=H_1$  and a red colour line for ${\rm DM}=A_1$. These correspond to scenario B with 
the specific choices  $\Delta_n =m_{A_2}-m_{H_1} =70$ GeV and  $\Delta_{n'}=m_{H_2}-m_{A_1} =50$ GeV.
}
  \label{deltar-ILC}
  \end{center}
\end{figure}

\begin{figure}
\begin{center}
    \includegraphics[scale=0.3]{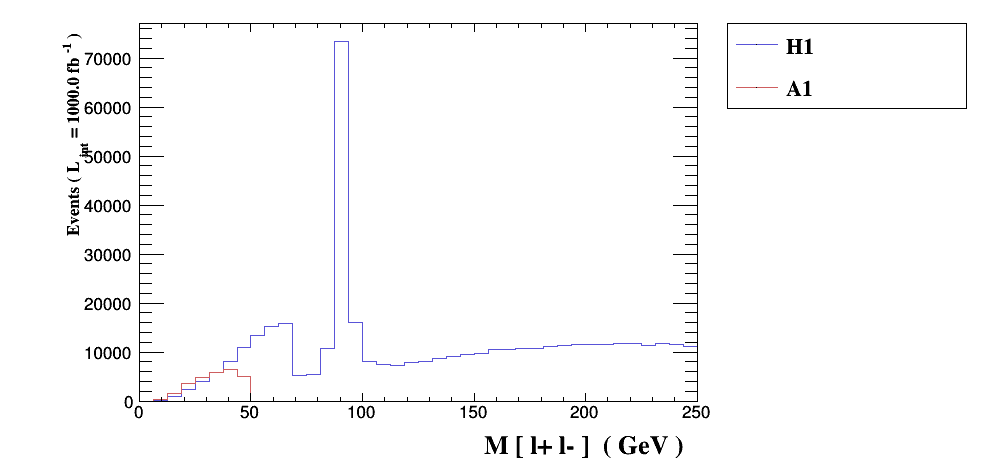}
\caption{Spectra in  invariant mass of the leptons for the processes $e^+e^-\to 2 l + 2 {\rm DM}$, where the distributions are identified by  a  blue colour line for  ${\rm DM}=H_1$  and a red colour line for ${\rm DM}=A_1$. These correspond to scenario B with 
the specific choices  $\Delta_n =m_{A_2}-m_{H_1} =70$ GeV and  $\Delta_{n'}=m_{H_2}-m_{A_1} =50$ GeV.
}
  \label{mll-1-ILC}
  \end{center}
\end{figure}

\begin{figure}
\begin{center}
      \includegraphics[scale=0.3]{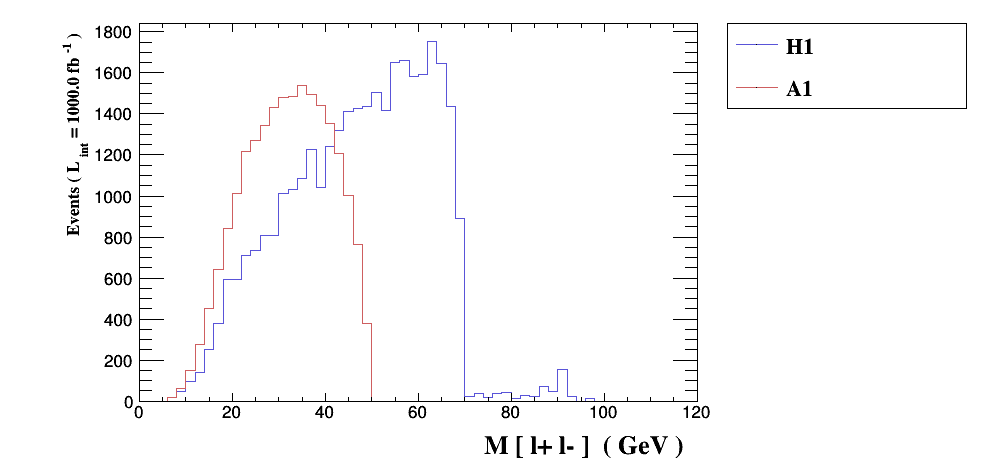}
\caption{Spectra in  invariant mass of the leptons for the processes $e^+e^-\to 2 l + 2 {\rm DM}$, where the distributions are identified by  a  blue colour line for  ${\rm DM}=H_1$  and a red colour line for ${\rm DM}=A_1$. These correspond to scenario B with 
the specific choices  $\Delta_n =m_{A_2}-m_{H_1} =70$ GeV and  $\Delta_{n'}=m_{H_2}-m_{A_1} =50$ GeV. Here, we include the additional cuts
in $\cancel{E}_T$  and $\Delta R(l^+,l^-)$.
}
  \label{mll-2-ILC}
  \end{center}
\end{figure}

\begin{figure}
\begin{center}
    \includegraphics[scale=0.3]{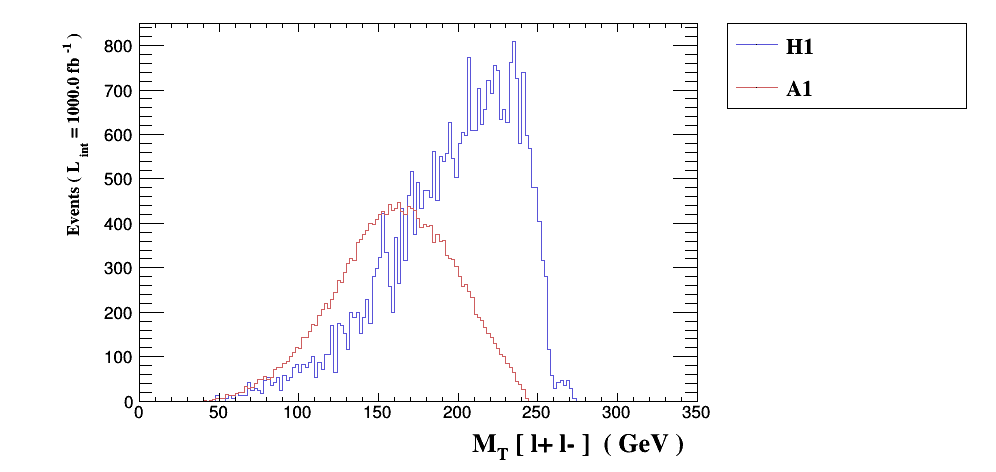}\\
\caption{Spectra in transverse mass of the final state for the processes $e^+e^- \to 2 l + 2 {\rm DM}$, where the distributions are identified by  a  blue colour line for  ${\rm DM}=H_1$  and a red colour line for ${\rm DM}=A_1$. These correspond to scenario B with 
the specific choices  $\Delta_n =m_{A_2}-m_{H_1} =70$ GeV and  $\Delta_{n'}=m_{H_2}-m_{A_1} =50$ GeV.
Here, we include the additional cuts
in $\cancel{E}_T$  and $\Delta R(l^+,l^-)$.
}
  \label{mT-ILC}
  \end{center}
\end{figure}

{
We here analyse  the signature $  2 l +  
 \cancel{E}_T$ at a future electron-positron machine, which is  induced by the processes $e^+ e^- \to 2 l + 2 H_1 (2 A_1)$. 
In order to demonstrate how to probe our scenario across two different collider environments,
we utilise the same model configuration as in the LHC analysis of the previous section, so we select scenario B with the same choice of 
$\Delta_n$ and $\Delta_{n'}$ (for the kinematical analysis).
}
As possible energies of a future
$e^+e^-$ collider, we adopt $\sqrt s_{ee}=250$, 350, 500 and 1000 GeV, with     $L=1000$ fb$^{-1}$ in all cases. 
We also assume the following beam polarisations:  80\% for the $e^-$ beam and 30\%  for the $e^+$ beam, though neither of these is necessary to uphold our forthcoming
conclusions. Cuts are the same as in the LHC case (initially). Again, as it happened for the latter, the cross-sections at an $e^+e^-$ machine   depend on the mass splitting $m_{A_2}  -m_{H_1}$  for the  $e^+ e^- \to 2 l + 2 H_1 $ channel and on  $m_{H_2}  -m_{A_1}$ for the channel $e^+ e^- \to 2 l + 2 A_1$. Indeed, at a future $e^+e^-$ machine, the detector resolution is even better than
at the LHC, so we expect to be able to see the two DM components of our $Z_3$ symmetric I(2+1)HDM scenario even more strikingly. 
To start with,  Fig. \ref{N1} illustrates that,
at the inclusive level, production rates of the above two processes are extremely significant, so the conditions are now much more favourable than the hadron machine. (Here, we have shown rates for the $H_1$ case, but they are overall similar for the $A_1$ one.)  
Furthermore, Fig. \ref{Deltan-ILC} illustrates that no matter the value of $e^+e^-$ collider energy, our choice of the BP within scenario B complies only somewhat with the aforementioned conditions for the total and relative rates of the cross-sections corresponding to the DM candidates while  Tab. \ref{sigmah1a1-ILC} shows  
their actual values at $\sqrt{s} =500$ GeV, the energy configuration that we adopt for the forthcoming kinematic analysis. 
By investigating the usual distributions in $\cancel{E}_T$, $p_T(l)$, $\eta(l)$ and $\Delta R(l^+,l^-)$, see Figs.~\ref{met-ILC-1}--\ref{deltar-ILC},
it is clear that their shapes are rather different for the
two Hermaphrodite DM contributions, owing to the fact that the collider energy chosen is more comparable to the inert (pseudo)scalar masses than that of the
LHC. However, they are such that the $A_1$ component is always subleading with respect to the $H_1$ one. In fact, even the $M(l^+l^-)$ distribution, Fig.
\ref{mll-1-ILC}, where the two contributions are very different (primarily because of a large on-shell $Z$ contribution in the $H_1$ case which is absent in the $A_1$ one), would not afford one to separate them.
\begin{table}[htp]
	\caption{Cross-section  for the processes $e^+e^- \to 2 l + 2 {\rm DM}$, with ${\rm DM} = H_1, A_1$, taking $L=1000$ fb$^{-1}$ as well as $m_{H_1} = 53 $ GeV, $m_{A_1} = 103 $ GeV,  $m_{A_2} = 123 $ GeV, $m_{H_2} = 153 $ GeV, $g_1=0.029$ with other parameters as in scenario B.}
	\label{sigmah1a1-ILC}
\begin{center}
\begin{tabular}{|c|c|c|c|}
\hline
DM & $\sigma (e^+ e^- \to 2 l + 2 {\rm DM}) $ & Event rates \\ \hline
$H_1 $ & 0.586 pb & $5.86 \times 10^{5}$ \\ \hline
$A_1 $ & 0.027 pb & $2.7 \times 10^{4}$ \\ \hline
\end{tabular}
\end{center}
\end{table}
Therefore, it is  appropriate to enforce some additional cuts, in order to make the two cross-sections for $H_1$ and $A_1$ more comparable.
By inspecting Figs. \ref{met-ILC-1}--\ref{deltar-ILC}, we choose the phase space regions where $ \cancel{E}_T< 120$ GeV and  $\Delta R (l, l) < 1.4$.
In the presence of these additional selections, it becomes clear that the two Hermaphrodite DM contributions become similar in normalisation while retaining 
sufficiently different shapes so to allow one to attempt their extraction, in both the invariant mass spectrum of the di-lepton pair (Fig.~\ref{mll-2-ILC}) and transverse mass of the 
final state (Fig.~\ref{mT-ILC}).

\section{Conclusions}

In this paper,  we have studied  a realisation of the 3HDM, wherein  one doublet is active, and two are inert  
(hence it is termed I(2+1)HDM), which, in the presence of a softly broken $Z_3$
symmetry, yields two DM candidates, in the form of the lightest CP-even and 
CP-odd states from the inert sector, $H_1$ and $A_1$, respectively. These two states, emerging from the same sector (hence with the same $Z_3$ properties), while having opposite CP quantum numbers, are not mass degenerate and have different gauge couplings, so that they cannot be ascribed to being the real and imaginary part of a single complex field. Therefore, they have been called Hermaphrodite DM. In the presence of constraints coming from the theoretical and experimental side, we have been able to isolate an expanse of I(2+1)HDM parameter space over which 
the two states $H_1$ and   $A_1$ are at the EW scale with a mass separation of order 50 GeV.  As the next-to-lightest CP-odd and CP-even states from the inert
sector, $A_2$ and
$H_2$, respectively, can decay into the DM candidates via a $Z$ boson, we have pursued here some possible signals of such Hermaphrodite DM, consisting
of $2l + \cancel{E}_T$  final states, which can be produced at both the LHC and a  future electron-positron collider. In the first case, the hard  process involved is
$q\bar q\to Z\to 2l 2H_1(2A_1)$ whereas in the second case this is     $e^+e^- \to Z\to 2l 2H_1(2A_1)$. The fact that the two Hermaphrodite DM states have a common final state enables one to potentially see these simultaneously in differential distributions that would have a distinctive shape carrying the imprint of the
two underlying components, each corresponding to a different DM candidate, at both the hadron and lepton collider. We have proven this to be the case for several observables by
using a parton-level MC analysis, albeit without a signal-to-background study. Therefore,  we
encourage experimentalists to look into this I(2+1)HDM hallmark phenomenology, as it would manifest itself in one of the most studied final states at both the aforementioned machines. Altogether, we expect that some evidence of two-component Hermaphrodite DM could first be seen at the LHC and eventually be 
characterised at a future $e^+e^-$ collider.

\begin{acknowledgments}
SM acknowledges support from the STFC Consolidated Grant ST/L000296/1 and is partially financed through the NExT Institute.
TS is supported in part by the JSPS
KAKENHI Grant Number 20H00160. TS and SM are partially supported by the Kogakuin
University Grant for the project research ``Phenomenological study of new physics models
with extended Higgs sector''. JH-S acknowledges the support by SNI-CONACYT (M\'exico), VIEP- BUAP and PRODEP-SEP (M\'exico) under the grant ‘Higgs and Flavour Physics’. DH-O acknowledges the support by CONACYT (M\'exico) and VIEP- BUAP.
\end{acknowledgments}




\pagebreak
\section*{References}

\providecommand{\noopsort}[1]{}\providecommand{\singleletter}[1]{#1}%

\end{document}